\documentstyle[12pt,cite,psfig]{article}

\newlength{\titlesep}
\newlength{\authorsep}
\makeatletter

\renewcommand{\thesection}{\Roman{section}}

\def\fnum@figure{FIG.~\thefigure}

\newcounter{figureparent}
\@addtoreset{figure}{figureparent}

\@addtoreset{equation}{section}
\renewcommand{\theequation}{(\arabic{section}.\arabic{equation})}
\def\@eqnnum{{\rm \theequation}}

\newcounter{eqnparent}
\@addtoreset{equation}{eqnparent}

\renewcommand{\abstract}{\if@twocolumn
  \section*{Abstract}
  \else
  \begin{center}
    {\bf Abstract\vspace{-.5em}\vspace{0pt}}
  \end{center}
  \fi}
\renewcommand{\endabstract}{\if@twocolumn\else\endquotation\fi}

\renewcommand{\appendix}{\par
  \setcounter{section}{0}
  \setcounter{subsection}{0}
  \renewcommand{\thesection}{Appendix~\Alph{section}}
  \renewcommand{\theequation}{(\Alph{section}.\arabic{equation})}}

\newcommand{\thismonth}{\ifcase\month\or
 January\or February\or March\or April\or May\or June\or
 July\or August\or September\or October\or November\or December\fi
 \space \number\year}

\newcommand{\preprintnumber}[1]
{\begin{flushright}
  \begin{tabular}{l} #1 \end{tabular}
  \end{flushright}}

\makeatother

\newcommand{\gsim}%
{\mathrel{\mbox{\raisebox{-1.0ex}%
{$\stackrel{\textstyle >}{\textstyle \sim}$}}}}
\newcommand{\lsim}%
{\mathrel{\mbox{\raisebox{-1.0ex}%
{$\stackrel{\textstyle <}{\textstyle \sim}$}}}}
\newcommand{\bsl}{$b \rightarrow s\,\ell\, {\overline \ell}$}
\newcommand{\bsll}{$b \rightarrow s\,\ell^+\ell^-$}
\newcommand{\bsnn}{$b \rightarrow s\,\nu\, {\overline \nu}$}

\newcommand{\bsg}{$b\rightarrow s\,\gamma$}
\newcommand{\bb}{$B^0$-${\overline B}^0$}
\newcommand{\kk}{$K^0$-${\overline K}^0$}
\newcommand{\br}{{\rm B}}
\newcommand{\bsln}{\br $(b \rightarrow c\ e \ \overline{\nu})$}
\newcommand{\usm}{$m_0$}
\newcommand{\mgx}{$M_{gX}$}
\newcommand{\ax}{$A_X$}
\newcommand{\tb}{$\tan\beta$}
\newcommand{\cs}{$C_{7}(m_b)$}
\newcommand{\cn}{$C_{9}(m_b)$}
\newcommand{\ct}{$C_{10}(m_b)$}
\newcommand{\vev}[1]{\left\langle #1 \right\rangle}
\def\sign{\mathop{\rm sign}\nolimits}

\newcommand{\Journal}[4]{{#1} {\bf #2}, {#4} {(#3)}}
\newcommand{\pl}{\sl Phys.~Lett.}
\newcommand{\plb}{\sl Phys.~Lett.~{\bf B}}
\newcommand{\prp}{\sl Phys.~Rep.}
\newcommand{\pr}{\sl Phys.~Rev.}
\newcommand{\prd}{\sl Phys.~Rev.~{\bf D}}
\newcommand{\prl}{\sl Phys.~Rev.~Lett.}
\newcommand{\np}{\sl Nucl.~Phys.}
\newcommand{\npb}{\sl Nucl.~Phys.~{\bf B}}
\newcommand{\ptp}{\sl Prog.~Theor.~Phys.}

\newcommand{\zpc}{\sl Z.~Phys.~{\bf C}}

\newcommand{\ibid}{\it ibid.}

\begin{document}
\baselineskip 18pt

\begin{titlepage}
\preprintnumber{%
KEK-TH-483 \\
KEK Preprint 96-125\\
OU-HET 247 \\
TU-504 \\
September 28, 1996
}
\vspace*{\titlesep}
\begin{center}
{\LARGE\bf
\bsl\,
 in the minimal supergravity model
}
\\
\vspace*{\titlesep}
{\large $^1$Toru  Goto},\\
{\large $^2$Yasuhiro  Okada,}\\
{\large $^{1,2}$Yasuhiro  Shimizu},\\
and
{\large $^3$Minoru  Tanaka}\\
\vspace*{\authorsep}
{\it $^1~$Department of Physics, Tohoku University \\
  Sendai 980-77 Japan}
\\
\vspace*{\authorsep}
{\it $^2$Theory Group, KEK, Tsukuba, Ibaraki, 305 Japan }
\\
\vspace*{\authorsep}
{\it $^3$ Department of Physics, Osaka University \\
  Toyonaka, Osaka, 560 Japan }
\end{center}
\vspace*{\titlesep}
\begin{abstract}
The $b \rightarrow s\,\ell^+ \ell^-$ process is studied in the minimal
supergravity  model in detail.
Taking account of the long distance contributions from the 
$c\overline{c}$ resonances, we calculate the branching ratio and the
lepton forward-backward asymmetry in this model.
We find that there is a strong correlation between the branching ratios
of $b \rightarrow s\,\gamma$ and $b \rightarrow s\,\ell^+ \ell^-$
processes and
that the interference effect can change the
$b \rightarrow s\,\ell^+ \ell^-$ branching ratio in the off-resonance
regions by up to $\pm 15\%$ depending on the relative phase between the
long and short distance contributions.
Using various phenomenological constraints including the
branching ratio of $b \rightarrow s\,\gamma$, we show that there are
regions in the parameter space where the branching ratio of
$b \rightarrow s\,\ell^+ \ell^-$ is enhanced by about $50$\% compared
to the SM.
We also show that the branching ratio of 
$b \rightarrow s\,\nu{\overline \nu}$ is reduced at most by $10$\%
from the SM prediction.
\end{abstract}

\end{titlepage}


\section{Introduction}
\label{sec:intro}

Although the Standard Model (SM) of the elementary particle physics is
successful in explaining almost all experimental results,
it is possible that physics beyond the SM exists just above the
presently available energy scale. Since new physics  may affect
various processes at low energy such as the flavor changing neutral
current (FCNC) processes of $K$ mesons and $B$ mesons, new 
physics searches in these processes are as important as direct
particle searches at collider experiments. A prime example is the
\bsg\, process. Experimentally the inclusive branching 
ratio is determined as 
\br (\bsg) = $(2.32\pm0.57\pm0.35)\times 10^{-4}$ at the
CLEO experiment \cite{CLEO}. It is known that this process puts very strong
constraints on various new physics beyond the SM, for example two Higgs
doublet model and supersymmetric (SUSY) extension of the SM. Along with the 
\bsg\, process, another important rare $b$ decay
process is the \bsl\, decay. Although only
upper bounds on branching ratios are given by experiments for various
exclusive modes \cite{BSLL},
this process is expected to be observed in the near future at $B$
factories as well as at hadron machines.

In this paper we investigate the \bsl\,  decay
in the minimal supersymmetric standard model (MSSM), especially in the
minimal supergravity (SUGRA) model. The MSSM is now considered to be the 
most promising candidate beyond the SM. In the MSSM, SUSY partners such as
squarks, sleptons, higgsinos and gauginos can contribute to FCNC
processes through 
loop diagrams. In order to evaluate their contributions
quantitatively
it is necessary to specify how soft SUSY breaking
terms are generated. In particular, the soft SUSY breaking terms in the squark
sector become new sources of flavor mixing, and
the \kk\, mixing becomes too large if the squark
mixing is $O(1)$ and masses of SUSY partners are in below-TeV
region \cite{KKbar}.
In
the minimal SUGRA model it is assumed that the soft SUSY breaking terms 
are universal at the Planck or GUT scale. Flavor mixing at the
electroweak scale can be determined by solving the relevant
renormalization group equations (RGEs) from the Planck to the low energy 
scale. It is shown that the constraint from the \kk\,
mixing is easily satisfied in this framework since masses of the first-
two-generation squarks with the same quantum numbers remain highly
degenerate at the low energy \cite{KKbar}.
 The FCNC processes involving the third
generation quarks and squarks can receive sizable SUSY contributions
due to the large top Yukawa coupling constant. 
In particular the \bsg\, process has been intensively studied both in
low energy SUSY Standard Models and in the minimal SUGRA model 
\cite{Bertolini,bsgamma,GO}. It was observed that the SUSY loop effects
can interfere with the SM amplitude constructively or destructively
depending on the parameters on the model whereas the charged Higgs
contribution is always  constructive.
For the minimal SUGRA model the \bb\,
mixing and CP violating parameter in the \kk\, mixing, $\epsilon_K$,
has been also investigated in Ref.~\cite{epsilonK}.
In this paper we
consider the \bsl\, process 
in addition to the \bsg\, process to see possible 
implication on the model
by future experiments. We observe that the predicted
branching ratio of the \bsg\, process and that of the \bsll\, process
are strongly correlated and thus their measurements are useful
to distinguish the SUGRA model from other extensions of the SM.

The \bsll\, process in the SUGRA model was analyzed by
Bertolini {\it et al.} in Ref.~\cite{Bertolini}.
Recently this process was reconsidered taking account of the measured 
branching ratio of \bsg\, and the top quark mass in the context of the
low-energy SUSY models as well as the minimal SUGRA model
\cite{Ali,cho}.
In Ref.~\cite{Ali} it was noted that the 
\bsll\, process is capable to resolve two-fold
ambiguity which cannot be distinguished from the branching ratio of 
\bsg.
In Ref.~\cite{cho} a more detailed analysis has been done in the minimal
SUGRA model.
Compared to them, our
calculation is improved in several points such as (1) numerically
solving RGEs with whole Yukawa matrices and soft SUSY breaking
parameters with the flavor  
mixing, (2) taking account of one-loop corrections in the Higgs effective
potential \cite{Vloop} in order to determine the proper vacuum
expectation value through the radiative electroweak symmetry breaking
scenario \cite{REWSB}, and (3) including the interference effect with the 
long distance contribution in calculating the lepton invariant mass
spectrum of the \bsll\, branching ratio.
It turns out that the third effect
can change the branching ratio in the off-resonance region by $\sim
\pm15$\% depending on the relative phase between the long and short
distance contributions.
Taking account of various phenomenological constraints including the
branching ratio of $b \rightarrow s\,\gamma$, we show that there are
regions in the parameter space where the branching ratio of
$b \rightarrow s\,\ell^+ \ell^-$ is enhanced by about  $50$\% compared
to the SM.
We also calculate the branching ratio of the 
\bsnn\, process in the minimal SUGRA model
and show that the branching ratio is reduced at most by $10$\%
from the SM value.

This paper is organized as follows. In section \ref{sec:model}, we
introduce the minimal SUGRA model and explain new sources of flavor
changing in this model.
 In section \ref{sec:bsll}, formulas for \bsll\,
decay are given including SUSY contributions.
In section
\ref{sec:results}, we present numerical results of our analysis.
In section \ref{sec:bsnn}, \bsnn\,
decay is discussed. 
Section \ref{sec:conclusion} gives  conclusions and discussions.
Various formulas are summarized in Appendices.


\section{Minimal SUGRA model}
\label{sec:model}

The MSSM Lagrangian is specified by the superpotential and the soft SUSY 
breaking terms.
The superpotential of the MSSM is given by
\begin{eqnarray}
  \label{superpotential}
  W_F =  \epsilon_{\alpha\beta} \left[
          f_{Uij}Q_i^{\alpha} H_2^\beta U_{j}
        + f_{Dij}H_1^\alpha Q_i^{\beta}D_{j}
        + f_{Eij}H_1^{\alpha}L_i^{\beta}E_j
        - \mu H_1^\alpha H_2^\beta \right],
\end{eqnarray}
where the chiral superfields $Q,U,D,L,E,H_1$ and $H_2$ transforms under 
$SU(3)_C \times SU(2)_L \times U(1)_Y$ as following representations:
\begin{eqnarray}
  \label{superfield}
  &&Q_i^\alpha = (3,2,\frac{1}{6}),~~
  U_i=(\overline{3},1,-\frac{2}{3}),~~
  D_i=(\overline{3},1,\frac{1}{3}),\\ \nonumber
  &&L_i^\alpha=(1,2,-\frac{1}{2}),~~
  E_i=(1,1,1),\\ \nonumber
  &&H_1^\alpha=(1,2,-\frac{1}{2}),~~
  H_2^\alpha=(1,2,\frac{1}{2}).
\end{eqnarray}
The suffices $\alpha,\beta=1,2$ are $SU(2)$ indices and 
$i,j=1,2,3$ are generation indices.
$\epsilon_{\alpha\beta}$ is the anti-symmetric
tensor with $\epsilon_{12}=1$.
A general form of the soft SUSY breaking terms is given by
\begin{eqnarray}
  \label{soft}
  -{\cal L}_{\rm soft} = &&
    (m_Q^2)_{ij} ~{\tilde q}^\dagger_{Li} ~{\tilde q}_{Lj}
  + (m_U^2)_{ij} ~{\tilde u}_{Ri}^{\ast} ~{\tilde u}_{Rj}
  + (m_D^2)_{ij} ~{\tilde d}_{Ri}^{\ast} ~{\tilde d}_{Rj}
\\ \nonumber &&
  + (m_L^2)_{ij} ~{\tilde \ell}^\dagger_{Li} ~{\tilde \ell}_{Lj}
  + (m_E^2)_{ij} ~{\tilde e}_{Ri}^{\ast} ~{\tilde e}_{Rj}
\\ \nonumber &&
  + \Delta^2_1 ~ h^\dagger_1 h_1 
  + \Delta^2_2 ~ h^\dagger_2 h_2
  + \epsilon_{\alpha\beta}\,( B\, \mu \,h_1^\alpha\, h_2^\beta + {\rm H.c.} )
\\ \nonumber &&
  + \epsilon_{\alpha\beta}\,
    (A_{Uij}\, {\tilde q}_{Li}^\alpha\,  h_2^\beta\, {\tilde u}^\ast_{Rj}
  +  A_{Dij}\,  h_1^\alpha\, {\tilde q}_{Li}^\beta\, {\tilde d}^\ast_{Rj}
  +  A_{Eij}\,  h_1^\alpha\, {\tilde \ell}_{Li}^\beta\, {\tilde e}^\ast_{Rj} +
  {\rm H.c.})
\\ \nonumber &&
  + (\frac{1}{2}m_{\tilde B}~{\tilde B}~{\tilde B}
  +  \frac{1}{2}m_{\tilde W}~{\tilde W}~{\tilde W}
  +  \frac{1}{2}m_{\tilde G}~{\tilde G}~{\tilde G} + {\rm H.c.}),
\end{eqnarray}
where ${\tilde q}_{Li}$, ${\tilde u}_{Ri}^\ast$, ${\tilde d}_{Ri}^\ast$,
${\tilde \ell}_{Li}$, ${\tilde  e}_{Ri}^\ast$, $h_1$ and $h_2$  
are scalar components of chiral superfields
$Q_i$, $U_i$, $D_i$,
$L_i$, $E_i$, $H_1$ and $H_2$ respectively, and ${\tilde B}$,
${\tilde W}$ and ${\tilde G}$ are $U(1)_Y$, $SU(2)_L$ and $SU(3)_C$ gauge
fermions.

In the minimal SUGRA model the soft SUSY breaking terms are assumed to
take the following universal structures at the GUT scale.
\begin{eqnarray}
  \label{eq:GUT}
  &&(m_Q^2)_{ij} = (m_U^2)_{ij} = (m_D^2)_{ij} = m_0^2~\delta_{ij},
\\ \nonumber
  &&(m_L^2)_{ij} = (m_E^2)_{ij} =  m_0^2~\delta_{ij},
\\ \nonumber
  && \Delta_1^2 = \Delta_2^2 = m_0^2,
\\ \nonumber
  && A_{Uij} = f_{Uij}~A_X,~
     A_{Dij} = f_{Dij}~A_X,~
     A_{Eij} = f_{Eij}~A_X,~
\\ \nonumber
  && m_{\tilde B} = m_{\tilde W} = m_{\tilde G} = M_{gX}.
\end{eqnarray}
With this initial condition, soft SUSY breaking parameters at 
the electroweak scale are calculated by solving the RGEs of the MSSM
\cite{RGE}.
We first solve the one-loop RGEs for the gauge coupling constants taking 
$\alpha_i(m_Z)$ as the input and determine the GUT scale, $M_{\rm GUT}$,
where the gauge couplings meet.
The Yukawa coupling constants at $M_{\rm GUT}$ are also calculated
by solving the RGEs from $m_Z$ to $M_{\rm GUT}$.
The values of the Yukawa coupling constants at the electroweak scale are
obtained by taking the quark/lepton masses, the
Cabbibo-Kobayashi-Maskawa (CKM)  matrix
elements and the $\tan\beta = \vev{h_2^0}/\vev{h_1^0}$ as input
parameters.
Then we solve the RGEs for all MSSM parameters downward with the GUT
scale boundary conditions \ref{eq:GUT} for each set of the universal
soft SUSY breaking parameters $(m_0,\, A_X,\, M_{gX})$.
We include all generation mixings in the RGEs for both Yukawa coupling
constants and the soft SUSY breaking parameters.
Next, we evaluate the Higgs potential at the electroweak scale and
require that the minimum of the potential gives a correct vacuum
expectation values of the neutral Higgs fields as
$\langle h_1^0 \rangle = v\cos\beta$ and
$\langle h_2^0 \rangle = v\sin\beta$ where $v = 174$ GeV.
This is known as the radiative electroweak symmetry breaking scenario
\cite{REWSB}.
The effective potential of neutral Higgs fields at the
electroweak scale is given by
\begin{eqnarray}
  \label{potential}
  V(h_1^0,h_2^0) &=&  (\mu^2+ \Delta_1^2)~|h_1^0|^2
  + (\mu^2+ \Delta_2^2)~|h_2^0|^2
  + (B\,\mu \,h_1^0\, h_2^0 + {\rm H.c.})
\\ \nonumber
  &&+ \frac{g^2+g'^2}{8}(|h_1^0|^2-|h_2^0|^2)^2 
  + V_{\rm loop},
\end{eqnarray}
where $V_{\rm loop}$ is the one-loop correction induced by the third
generation fermions and sfermions \cite{Vloop}.
The requirement of the radiative electroweak symmetry breaking
determines the magnitude of the SUSY Higgs mass parameter $\mu$ and the
soft SUSY breaking parameter $B$.
The explicit forms of $V_{\rm loop}$ and the condition of the radiative
electroweak symmetry breaking used in the present analysis are
given in \ref{sec:Vloop}.
At this stage, all MSSM parameters at the electroweak scale are
determined as functions of the input parameters
$(\tan\beta,\, m_0,\, A_X,\, M_{gX},\, \sign(\mu))$.

With use of the low energy SUSY parameters determined by the procedure
described above, we can calculate all the SUSY particle masses and the
mixing parameters.
The $6 \times 6$ mass matrix of up-type squark is written 
by
\begin{eqnarray}
  \label{up-type-mass}
  -{\cal L} 
  &=& ( {\tilde u}_L^\ast ,{\tilde u}_R^\ast)~M_{\tilde u}^2~
  \left(\begin{array}{c}
      {\tilde u}_L \\
      {\tilde u}_R
    \end{array}\right),
\\ \nonumber
  &=& ( {\tilde u}_{Li}^{\ast},{\tilde u}_{Ri}^{\ast})~
  \left(\begin{array}{cc}
      (m^2_{LL})_{ij}  & (m^2_{LR})_{ij} \\
      (m^2_{RL})_{ij} & (m^2_{RR})_{ij}
    \end{array}\right)~
  \left(\begin{array}{c}
      {\tilde u}_{Lj}\\
      {\tilde u}_{Rj}
    \end{array}\right),
\\
  (m^2_{LL})_{ij} 
  &=& (M_U^\dagger M_U)_{ij} + (m_Q^2)_{ij} + m_Z^2 \cos(2\beta)
      (\frac{1}{2}-\frac{2}{3}\sin^2 \theta_W) \delta_{ij},
\\
  (m^2_{RR})_{ij} 
  &=& (M_U M_U^\dagger )_{ij} + (m_U^2)_{ij} + m_Z^2 \cos(2\beta)
      (\frac{2}{3}\sin^2 \theta_W) \delta_{ij},
\\
  (m^2_{LR})_{ij}
  &=& (m^{2\dagger}_{RL})_{ij}
   =  -\mu \cot \beta (M_U^\dagger)_{ij} + (A_U^{\dagger})_{ij} v \sin \beta,
\end{eqnarray}
where $M_U$ is the up-type quark mass matrix, {\it i.e.}
$M_{Uij} = f_{Uji} v \sin \beta$, and ${\tilde u}_L$ is the up-type
component of $SU(2)$ doublet ${\tilde q}$. In this weak eigenstate, the
mass matrix is  not diagonal at the electroweak scale. The physical mass
eigenstate is given by diagonalizing the mass matrix,
\begin{equation}
  \label{mixing}
  {\tilde u}_I = ({\tilde U_U})_I^J~ \left(
     \begin{array}{cr}
      {\tilde u}_{L} \\
      {\tilde u}_{R}
    \end{array}\right)_J,
\end{equation}
\begin{equation}
   {\tilde U}_U ~(M_{\tilde u}^2)~ {\tilde U}_U^\dagger
    = {\rm diagonal},
\end{equation}
where ${\tilde u}_I$ is the mass eigenstate. The unitary matrix
${\tilde U}_U$ induces new flavor mixing in the up-type squark sector.
In a similar manner, we define the mixing matrices ${\tilde U}_D$,
${\tilde U}_\ell$ for the down-type squark and the slepton sectors.


\section{\bsll\, decay}
\label{sec:bsll}
In this section, we describe the calculation of the branching ratio 
and the lepton forward-backward asymmetry for the
\bsll \ $(\ell=e,\mu)$
process in the minimal SUGRA model.
We first introduce the effective Hamiltonian which is relevant for
the \bsll\, process \cite{GRI},
\begin{eqnarray}
  \label{eq:effective}
  H_{\rm eff} =  \frac{4 G_F}{\sqrt{2}}\sum_{i=1}^{10}~C_i(Q)~O_i(Q),
\end{eqnarray}
where $Q$ is the renormalization point. 
For the calculation of the branching ratio, the following three
operators are important,
\begin{eqnarray}
  \label{op}
O_7 &=& \frac{e}{16 \pi^2} m_b
               (\bar{s}_{L \alpha} \sigma_{\mu \nu} b_{R \alpha})
                F^{\mu \nu}                                    ,\\
O_9 &=& \frac{e^2}{16 \pi^2} (\bar{s}_{L \alpha} \gamma_
                \mu b_{L \alpha}) (\bar{\ell} \gamma^\mu \ell)   ,\\
O_{10} &=& \frac{e^2}{16 \pi^2}(\bar{s}_{L \alpha} \gamma_\mu 
                b_{L \alpha})
               (\bar{\ell} \gamma^\mu \gamma_5 \ell)             .
\end{eqnarray}
The explicit forms of all the effective 
operators $O_i(Q)$ are given in \ref{sec:ap1}.
Throughout this paper we neglect the strange
quark mass.

The coefficients 
$C_1(m_W)$--$C_{10}(m_W)$ are determined by matching the full theory
with the effective theory at the renormalization point $Q=m_W$.
The coefficients $C_1(m_W)$--$C_6(m_W)$ are given by
\begin{eqnarray}
  \label{wilson}
  C_2(m_W)=-\lambda_t,~~~C_i(m_W)=0,~(i=1,3{\mbox{--}}6),
\end{eqnarray}
where $\lambda_t = V_{tb}V_{ts}^\ast$. Note that
there is no SUSY contribution to these values at the tree level.
The coefficients $C_7(m_W)$--$C_{10}(m_W)$ are generated by 
one loop diagrams. In order to determine these coefficients at the
$m_W$ scale, we need to calculate
photon penguin, $Z$ penguin and box diagrams taking account of new
contributions in addition to the SM diagrams.
There are four classes of new contributions in the SUSY model;
charged Higgs boson($H^-$)--up-type quark loop,
chargino (${\tilde \chi}^-$)--up-type squark loop,
gluino (${\tilde G}$)--down-type squark loop, 
and neutralino (${\tilde \chi}^0$)--down-type squark loop.
$C_7(m_W)$ is obtained by calculating the photon penguin diagram
Fig.~\ref{fig:SM}(a).
$C_8(m_W)$ is also obtained by calculating the gluon penguin diagram
Fig.~\ref{fig:SM}(b).
There are three types of diagrams which contribute to $C_9(m_W)$;
the photon  penguin diagram Fig.~\ref{fig:SM}(c), the $Z$ penguin
diagram Fig.~\ref{fig:SM}(c) and the box diagram Fig.~\ref{fig:SM}(d).
Since we neglect the lepton mass,
there is no charged Higgs  contribution to the box diagram.
$C_{10}(m_W)$ is induced by the $Z$ penguin diagram Fig.~\ref{fig:SM}(c) and
the box diagram Fig.~\ref{fig:SM}(d). 
The explicit form of each contribution is given in \ref{sec:CI} with use 
of the mixing matrices at each vertex defined in \ref{sec:feynman} and
Inami-Lim functions given in \ref{sec:function}.

In order to calculate the \bsll\, decay amplitude,
we need the effective Hamiltonian at the $m_b$ scale.
By solving the RGEs in the leading logarithmic approximation 
(LLA) of QCD, $C_i(m_b)$ can be related with $C_i(m_W)$ as given in
\ref{sec:QCD} \cite{LLA}. 
With this effective Hamiltonian at the $m_b$ scale, we can calculate 
various physical observables. 
Since the bottom quark is much heavier than the QCD energy scale,
we can calculate the inclusive decay width as a free bottom quark decay.
This procedure is justified as a leading order approximation of the heavy
quark expansion \cite{HQET}.
The \bsll\, branching ratio
is given by
\begin{eqnarray}
  \label{decayrate}
&&\frac{d\, \br (b \rightarrow s\,\ell^+\ell^-)}{d\,\hat{s}}
= \br (b \rightarrow c\ e\ \overline{\nu})
 \frac{\alpha^2}{4 \pi^2}\left|\frac{\lambda_t}{V_{cb}}\right|^2
 \frac{1}{f_{\rm ph}(m_c/m_b)}w (\hat{s})
 \sqrt{1-\frac{4m_\ell^2}{s}}~~~~
\\
&&~~~~~~\times \left[ \vphantom{\frac{1}{1}}
 |C_9 + Y(\hat{s})|^2  \alpha_1 (\hat{s},\hat{m}_s,\hat{m}_\ell)
 + |C_{10}|^2 \alpha_2 (\hat{s},\hat{m}_s,\hat{m}_\ell)
\right. \nonumber \\ 
&&~~~~~~~~\left.
 + \frac{4}{\hat{s}} |C_7|^2 \alpha_3 (\hat{s},\hat{m}_s,\hat{m}_\ell)
+ 12 \alpha_4 (\hat{s},\hat{m}_s,\hat{m}_\ell) \mbox{Re}(C_7^\ast (C_9 
+ Y(\hat{s}))) \right], \nonumber
\end{eqnarray}
where $\hat{s} = (p_+ + p_-)^2/m_b^2$ and  $p_+(p_-)$ is the four momentum of
$\ell^+(\ell^-)$.
Here we normalize the branching ratio by the semileptonic 
branching ratio  \bsln\, in order to 
cancel the $m_b^5$ factor in the differential width.
The function $f_{\rm ph}(x)=1-8x^2+8x^6-x^8-24x^4\ln x$ is the phase space  
factor of the semileptonic decay width. Kinematical functions 
$\alpha_1$ -- $\alpha_4$ and $w(\hat{s})$ are given in \ref{sec:kfunc}.
As mentioned before we neglect the strange quark mass in the numerical
analysis. 
The lepton mass is, however, kept since lepton mass corrections are
important in the lower end of the lepton invariant mass spectrum.
By calculating the matrix elements of four quark operators
$O_1$--$O_6$ at the one loop level, we obtain $Y(\hat{s})$ 
\begin{eqnarray}
\label{eq:funcY}
Y(\hat{s}) 
&=& g(\frac{m_c}{m_b},\hat{s})
(3C_1 + C_2 + 3C_3 + C_4 + 3 C_5 + C_6) \\
&& - \frac{1}{2} g(1,\hat{s}) ( 4C_3 + 4C_4 + 3C_5 + C_6) 
- \frac{1}{2} g(0,\hat{s}) (C_3 + 3C_4) 
+ Y_{\rm res}(\hat{s}),
\nonumber
\end{eqnarray}
\begin{eqnarray}
\label{eq:funcg}
g(z,\hat{s})=
\left\{
\begin{array}{l}
 - \frac{4}{9} \ln z^2 + \frac{8}{27} 
                     + \frac{16z^2}{9\hat{s}}
\\
~~~~ - \frac{2}{9} 
\sqrt{1 - \frac{4 z^2}{\hat{s}}} \left( 2 + \frac{4 z^2}{\hat{s}} \right) 
\left( \ln \left| \frac{1 + \sqrt{1 - \frac{4 z^2}{\hat{s}}}}
                {1 - \sqrt{1 - \frac{4 z^2}{\hat{s}}}} \right|
              -i \pi \right)
\,\, \mbox{ for } \hat{s} > 4z^2,
 \\ 
- \frac{4}{9} \ln z^2 + \frac{8}{27} 
                     + \frac{16z^2}{9\hat{s}} \\ 
~~~~ - \frac{4}{9} 
\sqrt{ \frac{4 z^2}{\hat{s}}-1} \left( 2 + \frac{4 z^2}{\hat{s}} \right) 
\mbox{ arctan } \left( \frac{1} {\sqrt{\frac{4 z^2}{\hat{s}} - 1}} \right)
\,\, \mbox{ for } \hat{s} < 4z^2 .
\end{array}
\right.
\end{eqnarray}

In addition to these short distance contributions, there are long
distance contributions from the $c{\overline c}$ resonances,
$b \rightarrow s\, J/\psi \rightarrow \ell^+ \ell^-$ and
$b \rightarrow s\, \psi' \rightarrow \ell^+ \ell^-$.
Although we can avoid large contributions from these resonances by cutting 
the resonance regions of the lepton-invariant-mass spectrum, there can be 
sizable effects from interference between the short distance
contribution and the tail of the resonances. The resonance effects in 
the $b \rightarrow s$ transition have been investigated in connection to 
the long distance contribution of the \bsll\,
\cite{LMS,AMM,OAF}
as well as the \bsg\,
\cite{PDS} process.
In Eq.~\ref{eq:funcY} following Ref.~\cite{LMS,AMM} we have introduced the
resonance term $Y_{\rm res}(\hat{s})$,
\begin{eqnarray}
  \label{eq:res}
  Y_{\rm res}(\hat{s})
  &=& 
      \kappa \,\frac{3 \pi}{\alpha^2}  \sum_{i=J/\psi,\psi'}
     \frac{M_i \Gamma( i \rightarrow \ell^+\ell^-)/m_b^2}
     {\hat{s} - M_i^2/m_b^2 + i M_i \Gamma_i/m_b^2},
\end{eqnarray}
where $\kappa$ parametrizes the $b$--$s$--$J/\psi$ and
$b$--$s$--$\psi'$ couplings.
Its absolute value is determined from 
$\Gamma(b \rightarrow J/\psi X)$ and is 
given by $|\kappa| \sim 1$ \cite{LMS,AMM}.
In general $\kappa$ can have a non zero phase. In the following in order 
to see the effect of the phase, we show results with $\kappa = \pm 1$%
\footnote{In principle $\kappa$ can be different for $J/\psi$ and
  $\psi'$. But for simplicity, we have take the same value in
  Eq.~\ref{eq:res}.
 From the experimental data at least we can show that 
 the absolute value is almost the same both for $J/\psi$ and $\psi'$.}.
In the actual evaluation of the branching ratio the charm mass in
Eq.~\ref{eq:funcY} is taken to be the $D$ meson mass \cite{LMS}.
The choice of the charm mass is not important here since the branching
ratio depends on it very weakly.

Another observable which is expected to be measured with reasonable
accuracy in future experiments is the forward-backward asymmetry of the
lepton \cite{AMM}. 
In the center of mass frame of the lepton pair, this is defined as
\begin{eqnarray}
  \label{AFB}
  {\cal A}_{\rm FB} (\hat{s}) 
  &=& \frac{\int\limits_0^1 d(\cos\theta) \,
    \frac{d^2{\rm B}}{d(\cos\theta) \, d\hat{s}} 
    (b \to s\, \ell^+ \ell^-) - \int\limits_{-1}^0  d(\cos\theta)\,
    \frac{d^2{\rm B}}{d(\cos\theta) \, d\hat{s}}
    (b \to s\, \ell^+ \ell^-)}{\int\limits_0^1 d(\cos\theta) \,
    \frac{d^2{\rm B}}{d(\cos\theta) \, d\hat{s}}
    (b \to s\, \ell^+ \ell^-) + \int\limits_{-1}^0 d(\cos\theta) \,
    \frac{d^2{\rm B}}{d(\cos\theta) \, d\hat{s}} 
    (b \to s\, \ell^+ \ell^-)}~~~~~~~~~~
\\ \nonumber 
  &=&
- \frac{3 w(\hat{s}) 
           \sqrt{1-\frac{4\hat{m}_\ell^2}{\hat{s}}}
           C_{10} \left[ \hat{s} ( C_9 + \mbox{ Re } Y(\hat{s})) +
           2 C_7  \right] 
        }{
\left[ \vphantom{\frac{1}{1}}
 |C_9 + Y(\hat{s})|^2  \alpha_1  + C_{10}^2 \alpha_2
 + \frac{4}{\hat{s}} C_7^2 \alpha_3 + 12 \alpha_4 C_7 (C_9 
+ \mbox{ Re }Y(\hat{s})) \right]},
\end{eqnarray}
where $\theta$ is the angle between the momentum of the bottom quark and 
that of the $\ell^+$.


\section{Numerical results}
\label{sec:results}
As explained in Sec.~\ref{sec:model}, the MSSM parameters are
determined by solving the RGEs.
In the minimal SUGRA model, there are five 
SUSY parameters; the universal scalar mass \usm, the gaugino mass
\mgx, the universal $A$-parameter \ax, the SUSY invariant Higgs
mass $\mu$, the mixing parameter of Higgs boson $B$.
Using the condition that the electroweak symmetry is properly broken to
give the correct $Z$ boson mass, the theory contains four free
parameters, \tb, \usm, \mgx\, and \ax\, as well as the sign
of $\mu$.
We scan the parameters \usm, \mgx\, and \ax\, in the
range of \usm\, $\leq 2$ TeV, \mgx $\leq 2$ TeV and $|A_X|$ $\leq 5$ 
\usm\, for each fixed value of \tb. 
We also impose the following phenomenological constraints.
\begin{enumerate}
\item \bsg\, branching ratio:
\\
The branching ratio of the \bsg\, process is given by
\begin{eqnarray}
  \label{bsgamma}
  {\rm B}(b \rightarrow s \, \gamma) 
  = \frac{6\alpha}{\pi} \left| \frac{\lambda_t}{V_{cb}} \right|^2
    {\rm B}(b \rightarrow c\ e \ \overline{\nu}) |C_7(Q)|^2,
\end{eqnarray}
Most important theoretical ambiguity comes from the choice of
renormalization scale $Q$. The branching ratio changes by about
$\pm 30\%$ as the scale $Q$ is varied in the range of
$m_b/2 \leq Q \leq 2 m_b$.
We fix $Q = m_b$ in this analysis and discuss the ambiguity associated to
the QCD correction later.
From the measurement by CLEO \cite{CLEO},
the inclusive branch ratio is given by,
\begin{eqnarray}
  \label{br_bsgamma}
  1 \times 10^{-5} <  \br (b\rightarrow s\,\gamma) < 4.2 \times 10^{-5}.
\end{eqnarray}
\item From the recent experiment at LEP 1.5 \cite{LEP1.5}, we impose that all the
charged SUSY particles are heavier than 65 GeV.
\item All sneutrino masses are larger than 41 GeV \cite{PDG}.
\item The gluino mass is larger than 100 GeV.
The lower bound of the experimental gluino mass is given by Fermilab
TEVATRON collider \cite{TEV}. Since it depends on various SUSY parameters,
we take $100$ GeV as a conservative lower bound \nolinebreak%
\footnote{With use of the GUT relation of the gaugino masses
 Eq.~\ref{eq:GUT} and the LEP
1/1.5 constraints on charginos and neutralinos (above 2 and 5), a gluino
lighter than about 150 GeV is excluded for $\tan\beta \gsim 2$.
Therefore the precise value of the imposed gluino mass bound is not very
important. }.
\item From the neutralino search at LEP \cite{LEP}, we impose
\begin{equation}
    \label{znn}
    \Gamma(Z \rightarrow \chi \chi) < 8.4 \,{\rm MeV},
\end{equation}
\begin{equation}
    {\rm B}(Z \rightarrow \chi \chi' ),
    {\rm B}(Z \rightarrow \chi' \chi' )  < 2 \times 10^{-5},
\end{equation}
where $\chi$ is the lightest neutralino and $\chi'$ denotes other
neutralino.
\item The lightest SUSY particle is neutral.
\item The condition for not having a charge or color
  symmetry breaking vacuum \cite{aterm}.
\end{enumerate}

Throughout this paper we fix the top quark mass as $175$ GeV,
the bottom quark pole mass as 4.62 GeV and $\alpha_s(m_Z) = 0.116$.
In Fig.~\ref{fig:c7c9}, we show \cs, \cn\, and  \ct, each
of which is 
normalized to its SM value, for \tb\, = $3,30$.
In these figures we do not include the \bsg\, constraint.
We can see that in Fig.~\ref{fig:c7c9}
\cs\, can be quite different from the SM value and
even the opposite sign is allowed for \tb\, $=30$.
On the other hand, \cn\, and \ct\, differ  from the SM values by at most
5\% in the whole parameter space for both \tb\,$ =3,30$.
In the calculation of \cs, there is a one-loop diagram with internal 
stop and chargino which gives a large contribution when \tb\,
becomes large \cite{GO}.
When chargino has a sizable higgsino component, this
diagram is proportional to the product of the top and bottom Yukawa
coupling constants {\it i.e.} $\frac{m_t m_b}{\sin\beta\cos\beta}$
which grows as \tb\, when \tb\, is large. On the other hand, there are no 
such terms in the calculation of \cn\, and \ct. In fact, the
corresponding stop-higgsino diagram in \cn\, and \ct\, is proportional to
the square of the top Yukawa coupling constants, namely 
$\frac{m_t^2}{\sin^2\beta}$, which does not grow for large \tb.
Indeed \cn\, and \ct\, could be large if
\tb\, $\leq 1$, but within the framework of the minimal SUGRA model
\tb\, is only allowed to be larger than two as far as we require
that the top Yukawa coupling constant remains perturbative up to the GUT
scale.

We first show our numerical results for $b \rightarrow s\,\mu^+\mu^-$
and discuss the electron case later.
In Fig.~\ref{fig:spectrum} and Fig.~\ref{fig:FBA},
the branching ratio and the
forward-backward asymmetry in the SM are shown as functions of the
lepton pair invariant mass.
In the calculation of the \bsll\, branching ratio
we have used $m_c/m_b = 0.31$,
$|V_{cb}/\lambda_t| = 1.01$ and \bsln\, = 0.104 in Eq.~\ref{decayrate}.
We also show similar curves for 
the minimal SUGRA model with a particular set of 
parameters that \tb\, $=30$, \usm\, $=369$ GeV, \mgx\, $=100$ GeV, \ax\,
= \usm\, and  the sign of $\mu$ is positive.
This parameter set is chosen so that \cs\, has the same magnitude but
the opposite sign to the SM value.
We show the curves with $\kappa=\pm 1$ for both models. 
As can be seen in Fig.~\ref{fig:spectrum} and Fig.~\ref{fig:FBA},
there are large
contributions from the $J/\psi$ and $\psi'$ resonances. 
Since we are interested 
only in the short distance contribution, we consider the following 
two regions;
the low $s$ region, $4 m_\ell^2 < s < (m_{J/\psi}-\delta)^2$, and
the high $s$ region, $(m_{\psi'}+\delta)^2 < s < m_b^2$,
where  $\delta$ is introduced to cut the resonance regions and
we take $\delta = 100$ MeV here.
We can see that the sizable interference between the long and short
distance contributions even in these low and high $s$ regions.
At the asymmetric $B$ factory experiments, however, it may 
be possible to
determine the phase of $\kappa$ by measuring the lepton invariant
spectrum near the resonance regions. Therefore in the following, we
consider the branching ratio and asymmetry integrated in the above two
kinematical region with a choice of $\kappa = \pm 1$. These are defined
as
\begin{equation}
  \label{eq:intbr}
  {\rm B}^{\rm low(high)}
  = \int_{\rm low(high)}d\hat{s}~ {\rm B}(\hat{s}),
\end{equation}
\begin{equation}
  \label{eq:intafb}
  {\cal A}_{\rm FB}^{\rm low(high)}
  = \frac{\displaystyle{\int}_{\rm low(high)}d\hat{s}~
      \left(\int\limits_0^1 d(\cos\theta) \,
      \frac{d^2{\rm B}}{d(\cos\theta) \, d\hat{s}} 
       - \int\limits_{-1}^0  d(\cos\theta)\,
      \frac{d^2{\rm B}}{d(\cos\theta) \, d\hat{s}}
      \right)}
    {\displaystyle{\int}_{\rm low(high)}d\hat{s}~
      \left(\int\limits_0^1 d(\cos\theta) \,
      \frac{d^2{\rm B}}{d(\cos\theta) \, d\hat{s}}
       + \int\limits_{-1}^0 d(\cos\theta) \,
      \frac{d^2{\rm B}}{d(\cos\theta) \, d\hat{s}} 
      \right)}.
\end{equation}
Notice that for general phase of $\kappa$ the branching ratio takes the
value between the $\kappa = \pm 1$ cases.

In Fig.~\ref{fig:br30}, we show the
correlation between the branching ratios of the \bsg\, and
\bsll\, in the above two regions 
for \tb\, $=30$.
Fig.~\ref{fig:br30}(a) and (b) show the branching ratio of
\bsll\,
in the low $s$ region for $\kappa=\pm 1$, and 
Fig.~\ref{fig:br30}(c) and (d) correspond to the high $s$ region.
As already mentioned in connection with Fig.~\ref{fig:c7c9}, only \cs\,
can receive  sizable SUSY contributions. It is therefore clear from 
Eq.~\ref{bsgamma} and Eq.~\ref{decayrate} that the values of two
branching ratios lie on a parabola when
we neglect SUSY contribution to \cn\, and \ct.
This is seen in Fig.~\ref{fig:br30}.
If we take the experimental constraint on the branching ratio of 
\bsg\, into account, two separate regions are allowed.
One corresponds to the case when the sign of \cs\, is the same as that
of the SM, and the other corresponds to the case with the opposite sign.
 We can see
that the branching ratio of \bsll\, is enhanced about 50\%
in the latter case.
%
%
Although the branching ratio of \bsll\, in the
low $s$ region changes $\pm 15$\% depending on the sign of $\kappa$,
we can distinguish the sign of \cs\, from the branching ratio integrated 
in this region.
 On the other hand in the
high $s$ region, the branching ratio of
\bsll\,
depends on the sign of $\kappa$ significantly.

We also show  the correlations between the branching ratio of the 
\bsg\, and the forward-backward asymmetry of the
\bsll\,  in Fig.~\ref{fig:bsfb}.
Four figures correspond to the case $\kappa=\pm 1$ and low/high regions.
We can see that the asymmetry is also useful to distinguish the sign of 
\cs.

We vary the renormalization point $Q$ from $m_b/2$ to $2m_b$ in order
to study the renormalization point dependences, which are also shown
in Fig.~\ref{fig:br30} and Fig.~\ref{fig:bsfb}.
We see that the tendency that the branching ratio change along the
parabola.
This is because the change of the renormalization point
$Q$ mainly affects $C_7(Q)$.
This means that we can make a prediction of
the branching ratio of \bsll\, without much ambiguity 
as far as we use the experimental value of the \bsg\, branching ratio%
\footnote{It is important, however, to reduce the ambiguity of
  the renormalization point in order to put constraints on SUSY
  parameter space from the \bsg\, branching ratio.}.
Fig.~\ref{fig:chbr} shows that the branching ratio of
\bsll\, in the low $s$ region as a function
of the chargino mass and the light stop mass for $\kappa=1$ and 
\tb\, $=30$ taking account of the \bsg\,
constraint. The points where the branching ratio of the 
\bsll\, is enhanced about 50 \% compared to the SM
correspond to the case that the \cs\, has the opposite sign to the
SM. It is interesting to see such parameters correspond to relatively light 
SUSY particles ($m_{\chi^-} \lsim$ 130 GeV, $m_{\tilde t} \lsim$ 250 GeV)
but beyond the reach of LEP II. We have also analyzed the case of small \tb,
for example \tb = 3.
In this case  the \cs\, cannot change its sign as shown in 
Fig.~\ref{fig:c7c9}, thus the branching ratios change
within $\pm 5$\% after taking into account the \bsg\,
constraint.

We also calculated the branching ratio and the asymmetry for the
$b \rightarrow s\,e^+e^-$ process. The only difference from the 
$b \rightarrow s\,\mu^+\mu^-$ case is that the lower limit of the
lepton invariant mass becomes smaller. Since the $C_7(m_b)$ term gives
dominant contribution in the region near the kinematical lower limit,
the branching ratio and asymmetry integrated in the low $s$ region
change from those for
$b \rightarrow s\,\mu^+\mu^-$. Compared to Fig.~\ref{fig:br30}~(a),
for example, the $b \rightarrow s\,e^+e^-$ branching ratio is enhanced
by $\sim$5\% at B$(b \rightarrow s\,\gamma) = 1.0 \times 10^{-4}$ and 
$\sim$30\% at B$(b \rightarrow s\,\gamma) = 4.2 \times 10^{-4}$ for the SM
branch and by $\sim$5\% and $\sim$20\% respectively for the branch with
opposite 
sign of $C_7(m_b)$. It is worth while noting that we can distinguish  the
sign of $C_7(m_b)$ by looking at the low $s$ region  
 in the $b \rightarrow s\,e^+e^-$ mode just as 
in the $b \rightarrow s\,\mu^+\mu^-$ case. On the other hand the
branching ratio and asymmetry integrated in the high $s$ region do not
change noticeably from the $b \rightarrow s\,\mu^+\mu^-$ case.

Let us now compare our results with those in Ref.~\cite{cho}.
As explained in Sec.~\ref{sec:model}, we have included the one-loop
correction term $V_{\rm loop}$ in the Higgs potential to find
appropriate parameter sets.
This correction, however, mainly affects the mass of the lightest
neutral Higgs boson, which does not directly contribute to the FCNC processes.
Consequently the effect of this improvement is rather small%
\footnote{
  The effect on the lightest Higgs mass will be important in finding
  allowed SUSY parameter regions when the experimental bound of the
  lightest Higgs  mass is raised.
}.
The most important difference comes from the long distance contributions 
of the $c\overline{c}$ resonances.
In Ref.~\cite{cho} \bsll\, branching ratio is calculated with the
short distance contributions only, omitting the $J/\psi$ and $\psi'$
resonance regions from the integration range of the lepton pair
invariant mass.
We show that even if the resonance regions are avoided, the
interference effect between the short distance contributions and the
tail of the resonances gives $\sim \pm 15\%$ ambiguity to the \bsll\,
branching ratio since the detail of the $b$--$s$--$J/\psi$ and
$b$--$s$--$\psi'$ couplings, which are parametrized by the phase of
$\kappa$ in our present analysis, is theoretically unknown.
We see that this ambiguity is larger than the short distance effects
from the SUSY contributions unless the $C_7$ changes its sign.
Thus, it is difficult to extract information about the SUSY
parameters from the branching ratio and the forward-backward asymmetry
of \bsll\, without  knowledge of the long distance contributions.
This ambiguity will be reduced experimentally by the measurement of the 
behavior of the lepton pair invariant mass spectrum around the
resonances.
This may be achieved before the branching ratio in the
off-resonance regions is measured since a large number of events
is expected near the resonance regions.


\section{\bsnn\, decay}
\label{sec:bsnn}
In this section, we present the numerical result of the branching ratio
of \bsnn\, in the minimal SUGRA model.
For the calculation of this branching ratio, we need to introduce a new
operator to the effective Hamiltonian Eq.~\ref{eq:effective},
\begin{equation}
  \label{eq:O11}
O_{11} = 
             \frac{g^2}{16 \pi^2}(\bar{s}_{L \alpha} \gamma_\mu 
                b_{L \alpha}) \sum_{i=e,\mu,\tau}
               (\bar{\nu}_i \gamma^\mu (1-\gamma_5 )\nu_i).
\end{equation}
The corresponding Wilson coefficient $C_{11}$ is given in \ref{sec:CI}.
Note that $C_{11}$ does not receive QCD correction in LLA.
In addition to the SM contribution, there are the $Z$ penguin and the box
diagrams due to the charged Higgs boson and SUSY particles.
Since the effect of tau lepton mass in the loop diagram is small,
$C_{11}$ is calculated with 3 massless charged leptons.
Important difference from the \bsll\, process is that
no photon penguin diagram can contribute to the \bsnn\, process.
Thus SUSY contributions to $C_{11}$ are similar to those to $C_{10}$,
and no large SUSY contribution is induced.
The branching ratio of \bsnn\, is written as
\begin{eqnarray}
  \label{eq:brnn}
\sum_{i=e,\mu,\tau}{\rm B}(b\rightarrow s\, \nu_i \overline{\nu}_i)
=  3\ \br (b \rightarrow c\ e \ \overline{\nu})
        \frac{\alpha_W^2}{4 \pi^2}\left|\frac{\lambda_t}{V_{cb}}\right|^2
         \frac{1}{f_{\rm ph}(\frac{m_c}{m_b})}|C_{11}|^2.
\end{eqnarray}

In Fig.~\ref{fig:bsnunu} we show the scatter plot of the 
\bsnn\, branching ratio and the chargino
mass (the light stop mass). In this calculation we have taken into
account all constraints (1)--(7) in Sec.~\ref{sec:results}.
We see that the branching ratio does not exceed the SM value and
the change is at most $10$\%. 
This result does not depend much on the value of \tb.


\section{Conclusions}
\label{sec:conclusion}

In this paper, we have extensively examined the \bsl\, branching ratio
in the minimal SUGRA model.
By scanning the three-dimensional space of the soft SUSY breaking
parameters $m_0$, $M_{gX}$ and $A_X$ for various choices of $\tan\beta$, 
we have found that a parameter region, where the Wilson coefficient
$C_7(m_b)$ receives a large SUSY contribution, is still allowed under
the LEP 1.5 constraints provided that $\tan\beta$ is large.
On the other hand, the SUSY contributions to the coefficients $C_9(m_b)$
and $C_{10}(m_b)$ are much smaller than the SM contributions in the
whole allowed parameter space.
Consequently, there is a strong correlation between the predicted values
of the branching ratios of \bsg\, and \bsll.
Applying the measured bound of the \bsg\, branching ratio, we have shown 
that the predicted values of the \bsll\, branching ratio are separated
in two branches for a large $\tan\beta$:
one corresponds to the region where the sign of the \bsg\, amplitude is
the same as that in the SM and the other corresponds to the opposite
sign.
In the latter case, the \bsll\, branching ratio becomes at most
$\sim 50\%$ larger compared to the SM value.
The forward-backward asymmetry is also significantly different from the
SM in the same parameter region.
Since $m_{\chi^-} \gsim 100$ GeV is allowed for such a parameter region, 
it is possible to observe a large enhancement in the branching ratio of
\bsll\, even if no SUSY particle is found at LEP II.
We also calculated the branching ratio of \bsnn\, process and found that
it is reduced at most $10\%$ from the SM value.

There are several theoretical ambiguities in the calculation of \bsll\,
branching ratio, such as the $c \overline{c}$ resonance effect, the
renormalization point dependence, the strange quark mass and the higher
order corrections in the heavy quark expansion.
Among them we studied the renormalization point dependence and the
resonance effects in some detail.
The strange quark mass correction, which is of order $m_s^2/m_b^2$,
is estimated to be less important especially in the low $s$ region.
The higher order corrections in the heavy quark expansion is also
expected to be small \cite{HQET}.

The renormalization point dependence gives about $30\%$ ambiguity to
determine the magnitude of $C_7(m_W)$ from the measured value of \bsg\,
branching ratio.
This ambiguity, however, does not affect the correlation between the
branching ratios of \bsg\, and \bsll\, much.
We can make a rather definite prediction on the value of the \bsll\,
branching ratio with use of the measured \bsg\, branching ratio.

The $c \overline{c}$ resonance effect turns out to be important.
To deal with this effect, we have introduced a phenomenological
parameter $\kappa$ and have presented our results for $\kappa = \pm 1$
since the phase of $\kappa$ is not known theoretically.
We have pointed out that there are sizable ambiguities due to this
effect for both low and high $s$ regions.
This ambiguity will be reduced experimentally if the lepton invariant
mass spectrum near the resonances will be measured in some detail.
We have also found that in the low $s$ region the change of the \bsll\,
branching ratio due to the sign of $C_7(m_b)$ is larger than the
ambiguity induced from the phase of $\kappa$.
This fact enables us to distinguish the sign of $C_7(m_b)$ without the
knowledge of the phase of $\kappa$, by measuring the \bsll\, branching
ratio integrated in the low $s$ region.

\section*{Acknowledgments}

The authors would like to thank K.~Hikasa for carefully reading the
manuscript and giving useful comments
and S.~Uno for discussion on the possibility of the experimental
measurement on the lepton invariant mass spectrum near the resonance
regions.
The works of T.~G., Y.~O. and M.~T. were supported in part by the
Grant-in-Aid for Scientific Research from the Ministry of Education,
Science and Culture of Japan.

\newpage


\appendix

\section{One-loop Correction to the Higgs Potential}
\label{sec:Vloop}
The explicit form of $V_{\rm loop}$ in Eq.~\ref{potential} is given by
\begin{eqnarray}
  V_{\rm loop} &=& 
  \frac{3}{32\pi^2}
        \left[  m_{\tilde t_1}^4
                 \left(\ln\frac{m_{\tilde t_1}^2}{Q^2}-\frac{3}{2}\right)
              + m_{\tilde t_2}^4
                 \left(\ln\frac{m_{\tilde t_2}^2}{Q^2}-\frac{3}{2}\right)
              - m_t^4\left(\ln\frac{m_t^2}{Q^2}-\frac{3}{2}\right)
        \right]
\\ \nonumber
  &+&\frac{3}{32\pi^2}
        \left[  m_{\tilde b_1}^4
                 \left(\ln\frac{m_{\tilde b_1}^2}{Q^2}-\frac{3}{2}\right)
              + m_{\tilde b_2}^4
                 \left(\ln\frac{m_{\tilde b_2}^2}{Q^2}-\frac{3}{2}\right)
              - m_b^4\left(\ln\frac{m_b^2}{Q^2}-\frac{3}{2}\right)
        \right]
\\ \nonumber
  &+&\frac{1}{32\pi^2}
        \left[  m_{\tilde \tau_1}^4
                 \left(\ln\frac{m_{\tilde \tau_1}^2}{Q^2}-\frac{3}{2}\right)
              + m_{\tilde \tau_2}^4
                 \left(\ln\frac{m_{\tilde \tau_2}^2}{Q^2}-\frac{3}{2}\right)
              - m_\tau^4\left(\ln\frac{m_\tau^2}{Q^2}-\frac{3}{2}\right)
        \right],
\label{eq:Vloop}
\end{eqnarray}
where $Q$ denotes the renormalization point. The field-dependent masses are
given by
\begin{eqnarray}
  \label{mass}
  m_t = f_{U33}h_2^0,~~m_b = f_{D33}h_1^0,~~m_\tau = f_{E33}h_1^0,
\end{eqnarray}
\begin{eqnarray}
  \label{mass2}
  m^2_{\tilde t_{1(2)}} &=&
    \frac{1}{2}\left[
                2m_t^2+m^2_{Q33}+m^2_{U33}
                \mp\sqrt{(m^2_{Q33}-m^2_{U33})^2
                +4(-\mu f_{U33}h^0_1+A_{U33}h^0_2)^2}
               \right],~~~~~
\\
  m^2_{\tilde b_{1(2)}} &=&
    \frac{1}{2}\left[
                2m_b^2+m^2_{Q33}+m^2_{D33}
                \mp\sqrt{(m^2_{Q33}-m^2_{D33})^2
                +4(-\mu f_{D33}h^0_2+A_{D33}h^0_1)^2}
               \right],~~~~~
\\
  m^2_{\tilde \tau_{1(2)}} &=&
    \frac{1}{2}\left[
                2m_\tau^2+m^2_{L33}+m^2_{E33}
                \mp\sqrt{(m^2_{L33}-m^2_{E33})^2
                +4(-\mu f_{E33}h^0_2+A_{E33}h^0_1)^2}
               \right].~~~~~
\end{eqnarray}
For simplicity we neglect $D$-term contributions to the scalar masses.

The radiative electroweak symmetry breaking condition is
\begin{eqnarray}
\vev{\frac{\partial V}{\partial h_1^0}} &=&
\vev{\frac{\partial V}{\partial h_2^0}} ~=~0,
\label{eq:RB0}
\end{eqnarray}
where the bracket denotes the value at $h_1^0 = v\cos\beta$ and $h_2^0 =
v\sin\beta$.
We obtain the following equations for the SUSY Higgs mass parameter
$\mu$ and the soft SUSY breaking parameter $B$ from \ref{eq:RB0} with
the explicit form of the Higgs potential \ref{potential} and
\ref{eq:Vloop}:
\begin{eqnarray}
  \label{eq:RB}
  v^2 &=& \frac{4}{(g^2+g'^2)(\tan^2\beta-1)}
          \left(
            (\mu^2+\Delta_1^2)-(\mu^2+\Delta_2^2)\tan^2\beta
          \right.
\\ \nonumber
            &-&\frac{3f_{U33}^2}{16\pi^2}
            \left[
              (f(m^2_{\tilde t_1})+f(m^2_{\tilde t_2})-2f(m^2_t))\tan^2\beta
              +(A_{U33}^2\tan^2\beta-\mu^2)
               h(m^2_{\tilde t_1},m^2_{\tilde t_2})
            \right]
\\ \nonumber
            &+&\frac{3f_{D33}^2}{16\pi^2}
            \left[
              (f(m^2_{\tilde b_1})+f(m^2_{\tilde b_2})-2f(m^2_b))
              +(A_{D33}^2-\mu^2\tan^2\beta)
               h(m^2_{\tilde b_1},m^2_{\tilde b_2})
            \right]
\\ \nonumber
          &+&
          \left.
            \frac{f_{E33}^2}{16\pi^2}
            \left[
              (f(m^2_{\tilde \tau_1})+f(m^2_{\tilde \tau_2})-2f(m^2_\tau))
              +(A_{E33}^2-\mu^2\tan^2\beta)
               h(m^2_{\tilde \tau_1},m^2_{\tilde \tau_2})
            \right]
          \right),
\end{eqnarray}
\begin{eqnarray}
  \label{eq:RB2}
  &&\frac{-B\mu}{\sin\beta\cos\beta} = (\Delta_1^2 + \Delta_2^2 + 2\mu^2)
\\ \nonumber
            &&+\frac{3f_{U33}^2}{16\pi^2}
            \left[
              (f(m^2_{\tilde t_1})+f(m^2_{\tilde t_2})-2f(m^2_t))
              +(A_{U33}-\mu\cot\beta)(A_{U33}-\mu\tan\beta)
               h(m^2_{\tilde t_1},m^2_{\tilde t_2})
            \right]
\\ \nonumber
            &&+\frac{3f_{D33}^2}{16\pi^2}
            \left[
              (f(m^2_{\tilde b_1})+f(m^2_{\tilde b_2})-2f(m^2_b))
              +(A_{D33}-\mu\cot\beta)(A_{D33}-\mu\tan\beta)
               h(m^2_{\tilde b_1},m^2_{\tilde b_2})
            \right]
\\ \nonumber
            &&+\frac{f_{E33}^2}{16\pi^2}
            \left[
              (f(m^2_{\tilde \tau_1})+f(m^2_{\tilde \tau_2})-2f(m^2_\tau))
              +(A_{E33}-\mu\cot\beta)(A_{E33}-\mu\tan\beta)
               h(m^2_{\tilde \tau_1},m^2_{\tilde \tau_2})
            \right],
\end{eqnarray}
where $f(m^2) = m^2(\ln\frac{m^2}{Q^2}-1)$ and 
$h(m_1^2,m^2_2)= \frac{f(m^2_1)-f(m^2_2)}{m^2_1-m^2_2}$.

\section{The Effective Hamiltonian for \bsl}
\label{sec:ap1}

The effective Hamiltonian for the \bsl\, is given by
\begin{equation}
H_{\rm eff} =  \frac{4 G_F}{\sqrt{2}}\sum_{i=1}^{11} C_i (Q) O_i (Q) ,
\end{equation}
in which the operator basis is chosen to be
\begin{eqnarray}
O_1 &=& (\bar{s}_{L \alpha} \gamma_\mu b_{L \alpha})
               (\bar{c}_{L \beta} \gamma^\mu c_{L \beta}),     \\
O_2 &=& (\bar{s}_{L \alpha} \gamma_\mu b_{L \beta})
               (\bar{c}_{L \beta} \gamma^\mu c_{L \alpha}),     \\
O_3 &=& (\bar{s}_{L \alpha} \gamma_\mu b_{L \alpha})
               \sum_{q=u,d,s,c,b}
               (\bar{q}_{L \beta} \gamma^\mu q_{L \beta})  ,   \\
O_4 &=& (\bar{s}_{L \alpha} \gamma_\mu b_{L \beta})
                \sum_{q=u,d,s,c,b}
               (\bar{q}_{L \beta} \gamma^\mu q_{L \alpha})  ,   \\
O_5 &=& (\bar{s}_{L \alpha} \gamma_\mu b_{L \alpha})
               \sum_{q=u,d,s,c,b}
               (\bar{q}_{R \beta} \gamma^\mu q_{R \beta})    , \\
O_6 &=& (\bar{s}_{L \alpha} \gamma_\mu b_{L \beta})
                \sum_{q=u,d,s,c,b}
               (\bar{q}_{R \beta} \gamma^\mu q_{R \alpha})    , \\
O_7 &=& \frac{e}{16 \pi^2} m_b
               (\bar{s}_{L \alpha} \sigma_{\mu \nu} b_{R \alpha})
                F^{\mu \nu}                                    ,\\
O_8 &=& \frac{g_s}{16 \pi^2} m_b
(\bar{s}_{L \alpha} T_{\alpha \beta}^a \sigma_{\mu \nu} b_{R \alpha})
                G^{a \mu \nu}                                   , \\
O_9 &=& \frac{e^2}{16 \pi^2} (\bar{s}_{L \alpha} \gamma_
                \mu b_{L \alpha}) (\bar{\ell} \gamma^\mu \ell)   ,\\
O_{10} &=& \frac{e^2}{16 \pi^2}(\bar{s}_{L \alpha} \gamma_\mu 
                b_{L \alpha})
               (\bar{\ell} \gamma^\mu \gamma_5 \ell)             ,\\
O_{11} &=& 
               \frac{g^2}{16 \pi^2}(\bar{s}_{L \alpha} \gamma_\mu 
                b_{L \alpha}) \sum_{i=e,\mu ,\tau}
               (\bar{\nu}_i \gamma^\mu (1-\gamma_5 )\nu_i) .
\end{eqnarray}

\section{Wilson coefficients at the electroweak scale}
\label{sec:CI}
In this appendix, we give explicit forms of each contribution 
to Wilson coefficients at the electroweak scale~\cite{Bertolini,cho}.

\subsection{$C_7(m_W)$}
\label{sec:C7W}
\begin{eqnarray}
  \label{C7W}
  C_7(m_W) 
  &=& C_7^W + C_7^{H^-} + C_7^{{\tilde \chi}^-} 
  + C_7^{{\tilde G}}  + C_7^{{\tilde \chi}^0} ,
\\
  C_7^W 
  &=& \frac{3}{2}\lambda_t
  x_{tW}\left[\frac{2}{3}f_1(x_{tW})+f_2(x_{tW}) \right],
\\ 
  C_7^{H^-}
  &=& \frac{1}{2}\lambda_t x_{th}
  \left[ 
        \cot^2 \beta \left\{ \frac{2}{3}f_1(x_{th})+f_2(x_{th})\right\}
    +\left\{ \frac{2}{3}f_3(x_{th})+f_4(x_{th})\right\}
  \right],
\\
  C_7^{{\tilde \chi}^-}
  &=& -\sum_{\alpha=1}^2 \sum_{I=1}^{6}
    x_{W{\tilde u}_I}(\Gamma^{d\;\dagger}_{CL})_{\alpha 2}^I
    \left[
        (\Gamma^d_{CL})^{\alpha 3}_I
        \left\{f_1(x_{{\tilde \chi}^-_\alpha{\tilde u}_I})
            +\frac{2}{3}f_2(x_{{\tilde \chi}^-_\alpha{\tilde u}_I})
        \right\}
    \right.
\\ \nonumber
    &&~~~~~~~~ \left. +(\Gamma^d_{CR})^{\alpha 3}_I
        \frac{m_{{\tilde \chi}^-_\alpha}}{m_b}
        \left\{f_3(x_{{\tilde \chi}^-_\alpha{\tilde u}_I})
            +\frac{2}{3}f_4(x_{{\tilde \chi}^-_\alpha{\tilde u}_I})
        \right\}
   \right],
\\
  C_7^{{\tilde G}}
  &=&\frac{8}{9}\ \frac{g_s^2}{g^2}
    \sum_{I=1}^{6} x_{W{\tilde d}_I}(\Gamma^{d \;\dagger}_{GL})^I_2
    \left[(\Gamma^d_{GL})_I^3
          f_2(x_{{\tilde G}{\tilde d}_I})
          +(\Gamma^d_{GR})^3_I
          \frac{m_{\tilde G}}{m_b}f_4(x_{{\tilde G}{\tilde d}_I})
    \right],
\\
  C_7^{{\tilde \chi}^0}
  &=& \frac{1}{3} 
      \sum_{\alpha=1}^4 \sum_{I=1}^{6} x_{W{\tilde d}_I}
      (\Gamma^{d\;\dagger}_{NL})_{\alpha 2}^I
      \left[
        (\Gamma^d_{NL})^{\alpha 3}_I
        f_2(x_{{\tilde \chi}^0_\alpha{\tilde u}_I})
        +(\Gamma^d_{NR})^{\alpha 3}_I
        \frac{m_{{\tilde \chi}^0_\alpha}}{m_b}
        f_4(x_{{\tilde \chi}^0_\alpha{\tilde u}_I})
      \right],~~~~~~~~
\end{eqnarray}
where $x_{ij}= m^2_i/m^2_j$ and $m_i$ is the mass of the particle $i$.
$f_i(x)$ are the one loop functions, which are given in 
\ref{sec:function}. 
The matrix $\Gamma^d_{CL(R)}$ represents the coupling constant of
chargino--(up-type-)squark--(down-type-)quark,  $\Gamma^d_{NL(R)}$
represents that of neutralino--(down-type-)squark--(down-type-)quark 
and $\Gamma^d_{GL(R)}$
represents that of gluino--(down-type-)squark--(down-type-)quark,
which are found in \ref{sec:feynman}.

\subsection{$C_8(m_W)$}
\label{sec:C8W}

\begin{eqnarray}
  \label{C8W}
  C_8(m_W) 
  &=& C_8^W + C_8^{H^-} + C_8^{{\tilde \chi}^-} 
  + C_8^{{\tilde G}}  + C_8^{{\tilde \chi}^0},
\\
  C_8^W 
  &=& \frac{3}{2} \lambda_t  x_{tW}f_1(x_{tW}),
\\
  C_8^{H^-}
  &=&\frac{1}{2} \lambda_t x_{th}
  \left[ \cot^2\beta f_1(x_{th}) + f_3(x_{th}) \right],
\\
  C_8^{{\tilde \chi}^-}
  &=& -\sum_{\alpha=1}^2 \sum_{I=1}^{6}
     x_{W{\tilde u}_I}(\Gamma^{d\;\dagger}_{CL})_{\alpha 2}^I
    \left[
        (\Gamma^d_{CL})^{\alpha 3}_I
            f_2(x_{{\tilde \chi}^-_\alpha{\tilde u}_I})
            +(\Gamma^d_{CR})^{\alpha 3}_I
            \frac{m_{{\tilde \chi}^-_\alpha}}{m_b}
            f_4(x_{{\tilde \chi}^-_\alpha{\tilde u}_I})
    \right],~~~~~~~~
\\
  C_8^{{\tilde G}}
  &=&\frac{g_s^2}{g^2} 
    \sum_{I=1}^{6} x_{W{\tilde d}_I}
    (\Gamma^{d\;\dagger}_{GL})^I_2
    \left[ (\Gamma^d_{GL})^3_I
         \left\{ 3 f_1(x_{{\tilde G}{\tilde d}_I})
               +\frac{1}{3} f_2(x_{{\tilde G}{\tilde d}_I})
         \right\}
    \right.
\\ \nonumber &&~~~~~~~~
    \left.+(\Gamma^d_{GR})^3_I
          \frac{m_{\tilde G}}{m_b}
          \left\{3 f_3(x_{{\tilde G}{\tilde d}_I})
               +\frac{1}{3}f_4(x_{{\tilde G}{\tilde d}_I})
          \right\}
    \right],
\\
  C_8^{{\tilde \chi}^0}
&=& -  \sum_{\alpha=1}^4 \sum_{I=1}^{6}
       x_{W{\tilde d}_I} (\Gamma^{d\;\dagger}_{NL})_{\alpha 2}^I
    \left[
        (\Gamma^d_{NL})^{\alpha 3}_I
        f_2(x_{{\tilde \chi}^0_\alpha{\tilde d}_I})
        +(\Gamma^d_{NR})^{\alpha 3}_I
        \frac{m_{{\tilde \chi}^0_\alpha}}{m_b}
        f_4(x_{{\tilde \chi}^0_\alpha{\tilde d}_I})
    \right].~~~~~~~~
\end{eqnarray}

\subsection{$C_9(m_W)$}
\label{sec:C9W}

\begin{eqnarray}
  \label{C9W}
  C_9(m_W) &=& C_{9,\gamma} + C_{9,Z} + C_{9,box},
\\
  C_{9,\gamma} 
  &=& C_{9,\gamma}^W + C_{9,\gamma}^{H^-} 
    + C_{9,\gamma}^{{\tilde \chi}^-}
    + C_{9,\gamma}^{{\tilde G}}
    + C_{9,\gamma}^{{\tilde \chi}^0},
\\
  C_{9,Z} 
  &=& C_{9,Z}^W + C_{9,Z}^{H^-} 
    + C_{9,Z}^{{\tilde \chi}^-}
    + C_{9,Z}^{{\tilde G}}
    + C_{9,Z}^{{\tilde \chi}^0},
\\
  C_{9,box}
  &=& C_{9,box}^W + C_{9,box}^{{\tilde \chi}^-}
    + C_{9,box}^{{\tilde \chi}^0},
\\
  C_{9,\gamma}^{W}
  &=&\lambda_t
      \left[ x_{tW} 
             \left\{ \frac{2}{3}f_7(x_{tW})+f_8(x_{tW}) \right\}
             +\frac{4}{9}\left(\frac{\ln x_{tW}}{x_{tW}-1}-1 \right)
      \right],
\\
  C_{9,\gamma}^{H^-}
  &=& \lambda_t \cot^2\beta\;
      x_{th} \left[\frac{2}{3}f_5(x_{th}) - f_6(x_{th})\right],
\\
  C_{9,\gamma}^{{\tilde \chi}^-}
  &=& 2 \sum_{\alpha=1}^2 \sum_{I=1}^{6}
     x_{W{\tilde u}_I}
        (\Gamma^d_{CL})^{\alpha 3}_I
        (\Gamma^{d\;\dagger}_{CL})_{\alpha 2}^I
        \left[\frac{2}{3}f_6(x_{{\tilde \chi}^-_\alpha{\tilde u}_I})
        -f_5(x_{{\tilde \chi}^-_\alpha{\tilde u}_I})\right],
\\
  C_{9,\gamma}^{{\tilde G}}
  &=& - \frac{16}{9} \frac{g_s^2}{g^2}
      \sum_{I=1}^{6} x_{W{\tilde d}_I}
      (\Gamma^d_{GL})^3_I(\Gamma^{d\;\dagger}_{GL})^I_2
      f_6(x_{{\tilde G}{\tilde d}_I}),
\\
  C_{9,\gamma}^{{\tilde \chi}^0}
  &=& -\frac{2}{3} 
      \sum_{\alpha=1}^4 \sum_{I=1}^{6} x_{W{\tilde d}_I}
        (\Gamma^d_{NL})^{\alpha 3}_I
        (\Gamma^{d\;\dagger}_{NL})_{\alpha 2}^I
        f_6(x_{{\tilde \chi}^0_\alpha{\tilde d}_I}),
\\
  C_{9,Z}^{W}
  &=& - \left(-1+\frac{1}{4 \sin^2 \theta_W}\right)x_{tW}f_9(x_{tW}),
\\
  C_{9,Z}^{H^-}
  &=& \frac{1}{2} \left(-1+\frac{1}{4 \sin^2 \theta_W}\right) 
      \cot^2\beta  x_{tW} x_{th}
      \left[f_3(x_{th})+f_3(x_{th})\right],
\\ 
  C_{9,Z}^{{\tilde \chi}^-}
  &=&2
    \left(-1+\frac{1}{4 \sin^2 \theta_W}\right) 
    \sum_{\alpha,\beta=1}^2 \sum_{I,J=1}^{6}
        (\Gamma^d_{CL})^{\alpha 3}_I
        (\Gamma^{d\;\dagger}_{CL})_{\beta 2}^J
\\ \nonumber && \times
      \left[\delta_{\alpha\beta} 
        g_2(x_{{\tilde u}_J{\tilde \chi}^-_\beta},
            x_{{\tilde u}_I{\tilde \chi}^-_\alpha})
            \sum_{M=1}^{3}({\tilde U}_{U})^M_J
            ({\tilde U}^\dagger_{U})^I_M
      \right.
\\ \nonumber&&~~~~~~~~~
      \left.
        +\delta_{IJ}
        \left\{ 2 \sqrt{x_{{\tilde \chi}^-_\alpha{\tilde u}_I}
                        x_{{\tilde \chi}^-_\beta{\tilde u}_J}}
        g_1(x_{{\tilde \chi}^-_\alpha{\tilde u}_I},
            x_{{\tilde \chi}^-_\beta{\tilde u}_J})
        (U_-^\dagger)_\alpha^1(U_-)_1^\beta
       \right.
     \right.
\\&&~~~~~~~~
     \left.
       \left.
        +(\ln x_{W{\tilde u}_I}-
        g_2(x_{{\tilde \chi}^-_\alpha{\tilde u}_I},
            x_{{\tilde \chi}^-_\beta{\tilde u}_J})
        (U_+^\dagger)_\alpha^1(U_+)_1^\beta
      \right\}
     \right],
\\
  C_{9,Z}^{\tilde G}
  &=& 
     \frac{g_s^2}{g^2} \left(-1+\frac{1}{4 \sin^2 \theta_W}\right) 
     \frac{4}{3} \sum_{I,J=1}^{6} \sum_{M=1}^{3}
     (\Gamma^d_{GL})^6_I(\Gamma^{d\;\dagger}_{GL})^J_5
\\ \nonumber
   &&~~ \times (\Gamma^d_{GR})^M_J (\Gamma^{d\;\dagger}_{GR})^I_M
     g_2(x_{{\tilde d}_I{\tilde G}},x_{{\tilde d}_J{\tilde G}}),
\\
  C_{9,Z}^{{\tilde \chi}^0}
  &=& \frac{1}{2} 
      \left(-1+\frac{1}{4 \sin^2 \theta_W}\right) 
      \sum_{\alpha,\beta=1}^4 \sum_{I,J=1}^{6}
        (\Gamma^d_{NL})^{\alpha 3}_I
        (\Gamma^{d\;\dagger}_{NL})^J_{\beta 2}
\\ \nonumber && \times
      \left[\delta_{\alpha\beta} 
        g_2(x_{{\tilde d}_J{\tilde \chi}^0_\beta},
            x_{{\tilde d}_I{\tilde \chi}^0_\alpha})
            \sum_{M=1}^{3}({\tilde U}_{D})_J^M
            ({\tilde U}^\dagger_{D})^I_M
      \right.
\\ \nonumber&&~~~~~~~~~
      \left.
        +\delta_{IJ}
        \left\{- 2 \sqrt{x_{{\tilde \chi}^0_\beta{\tilde d}_I}
                        x_{{\tilde \chi}^0_\alpha{\tilde d}_I}}
        g_1(x_{{\tilde \chi}^0_\beta{\tilde d}_I},
            x_{{\tilde \chi}^0_\alpha{\tilde d}_I})
        G_{\beta\alpha}
       \right.
     \right.
\\ \nonumber &&~~~~~~~~
     \left.
       \left.
        +(\ln x_{W{\tilde d}_I}-
        g_2(x_{{\tilde \chi}^0_\beta{\tilde d}_I},
            x_{{\tilde \chi}^0_\alpha{\tilde d}_I})
        G_{\alpha\beta}
      \right\}
     \right],
\\
  C_{9,box}^{W}
  &=& \lambda_t \,
      \frac{1}{4\sin^2 \theta_W}\left[g_3(x_{tW},0)-g_3(0,0)\right],
\\ 
  C_{9,box}^{{\tilde \chi}^-}
  &=&\frac{1}{4\sin^2 \theta_W}
    \sum_{\alpha,\beta=1}^2 \sum_{I=1}^{6}\sum_{J=1}^{3}
        x_{W{\tilde \chi}_\alpha^-}
        (\Gamma^d_{CL})^{\alpha 3}_I
        (\Gamma^{d\;\dagger}_{CL})_{\beta 2}^I
\\ \nonumber
  &&~~~~~~~~~\times (\Gamma^l_{CL})^{\beta i}_J
        (\Gamma^{l\;\dagger}_{CL})_{\alpha  i}^J
        g_6(x_{{\tilde u}_I      {\tilde \chi}_\alpha^-},
           x_{{\tilde \nu}_J    {\tilde \chi}_\alpha^-},
           x_{{\tilde \chi}_\beta^- {\tilde \chi}_\alpha^-}),
\\
  C_{9,box}^{{\tilde \chi}^0}
  &=&\frac{1}{4\sin^2 \theta_W}
    \sum_{\alpha,\beta=1}^{4} \sum_{I,J=1}^{6}
        x_{W{\tilde \chi}_\alpha^0}
        (\Gamma^d_{NL})^{\alpha 3}_I
        (\Gamma^{d\;\dagger}_{NL})_{\beta 2}^I
\\\nonumber&&~~~~~
   \times \left[
      \left\{
        (\Gamma^\ell_{NL})^{\beta i}_J
        (\Gamma^{\ell\;\dagger}_{NL})_{\alpha  i}^J
       -(\Gamma^\ell_{NR})^{\alpha i}_J
        (\Gamma^{\ell\;\dagger}_{NR})_{\beta i}^J
      \right\} 
      g_6(x_{{\tilde d}_J{\tilde \chi}^0_\alpha}
          ,x_{{\tilde \ell}_I{\tilde \chi}^0_\alpha}
          ,x_{{\tilde \chi}^0_\beta {\tilde \chi}^0_\alpha})
    \right.
\\ \nonumber&&~~~~~
    -\left.
      \left\{
        (\Gamma^\ell_{NL})^{\beta i}_J
        (\Gamma^{\ell\;\dagger}_{NL})_{\alpha  i}^J
       -(\Gamma^\ell_{NR})^{\alpha i}_J
        (\Gamma^{\ell\;\dagger}_{NR})_{\beta  i}^J
      \right\} 
      2 \sqrt{x_{{\tilde \chi}^0_\beta {\tilde \chi}^0_\alpha}}
         g_5(x_{{\tilde d}_J{\tilde \chi}^0_\alpha}
          ,x_{{\tilde \ell}_I{\tilde \chi}^0_\alpha}
          ,x_{{\tilde \chi}^0_\alpha {\tilde \chi}^0_\alpha})
    \right].
\end{eqnarray}
The matrix $\Gamma^\ell_{CL(R)}$ represents the coupling constant of
chargino-sneutrino-lepton  and $\Gamma^\ell_{NL(R)}$
that of neutralino-slepton-lepton,
which are found in \ref{sec:feynman}.
 Note that index $i$ represents the 
generation of the final lepton and is not summed here.

\subsection{$C_{10}(m_W)$}
\label{sec:C10W}

\begin{eqnarray}
  \label{C10W}
  C_{10}(m_W) &=& C_{10,Z} + C_{10,box},
\\
  C_{10,Z}
  &=& C_{10,Z}^{W} + C_{10,Z}^{H^-} + C_{10,Z}^{{\tilde \chi}^-}
  + C_{10,Z}^{{\tilde G}} + C_{10,Z}^{{\tilde \chi}^-},
\\
  C_{10,box}
  &=& C_{10,box}^{W} + C_{10,box}^{{\tilde \chi}^-}
  + C_{10,box}^{{\tilde \chi}^0},
\\
  C_{10,Z}^i 
  &=& \frac{-\frac{1}{4\sin^2 \theta_W}}
           {-1+\frac{1}{4\sin^2 \theta_W}} C_{9,Z}^i
  ,~~~i=W,H^-,{\tilde \chi}^-,{\tilde G},{\tilde \chi}^0,
\\
  C_{10,box}^{W} 
  &=& -C_{9,box}^{i},~~~~~~~~~~i = W,{\tilde \chi}^-
\\
  C_{10,box}^{{\tilde \chi}^0}
  &=&\frac{1}{4\sin^2 \theta_W}
    \sum_{\alpha,\beta=1}^{4} \sum_{I,J=1}^{6}
        x_{W{\tilde \chi}_\alpha^0}
        (\Gamma^d_{NL})^{\alpha 3}_I
        (\Gamma^{d\;\dagger}_{NL})_{\beta 2}^I
\\\nonumber&&~~~~~
    \left[
     - \left\{
        (\Gamma^\ell_{NL})^{\beta i}_J
        (\Gamma^{\ell\;\dagger}_{NL})_{\alpha  i}^J
       +(\Gamma^\ell_{NR})^{\alpha i}_J
        (\Gamma^{\ell\;\dagger}_{NR})_{\beta i}^J
      \right\} 
      g_6(x_{{\tilde d}_J{\tilde \chi}^0_\alpha}
          ,x_{{\tilde \ell}_I{\tilde \chi}^0_\alpha}
          ,x_{{\tilde \chi}^0_\beta {\tilde \chi}^0_\alpha})
    \right.
\\ \nonumber&&~~~~~
    +\left.
      \left\{
        (\Gamma^\ell_{NL})^{\beta i}_J
        (\Gamma^{\ell\;\dagger}_{NL})_{\alpha  i}^J
       +(\Gamma^\ell_{NR})^{\alpha i}_J
        (\Gamma^{\ell\;\dagger}_{NR})_{\beta  i}^J
      \right\} 
      2 \sqrt{x_{{\tilde \chi}^0_\beta {\tilde \chi}^0_\alpha}}
         g_5(x_{{\tilde d}_J{\tilde \chi}^0_\alpha}
          ,x_{{\tilde \ell}_I{\tilde \chi}^0_\alpha}
          ,x_{{\tilde \chi}^0_\alpha {\tilde \chi}^0_\alpha})
    \right].
\end{eqnarray}

\subsection{$C_{11}(m_W)$}
\label{sec:C11W}

\begin{eqnarray}
  \label{C11W}
  C_{11}(m_W) &=& C_{11,Z} + C_{11,box},
\\
  C_{11,Z}
  &=& C_{11,Z}^{W} + C_{11,Z}^{H^-} + C_{11,Z}^{{\tilde \chi}^-}
  + C_{11,Z}^{{\tilde G}} + C_{11,Z}^{{\tilde \chi}^-},
\\
  C_{11,box}
  &=& C_{11,box}^{W} + C_{11,box}^{{\tilde \chi}^-}
  + C_{11,box}^{{\tilde \chi}^0},
\\
  C_{11,Z}^i 
  &=& \sin \theta^2_W C_{10,Z}
  ,~~~~~~~~~i=W,H^-,{\tilde \chi}^-,{\tilde G},{\tilde \chi}^0,
\\
  C_{11,box}^{W} 
  &=& 4 \sin\theta^2_W C_{10,box}^{W},
\\
  C_{11,box}^{{\tilde \chi}^-} 
  &=&-\frac{1}{4}
    \sum_{\alpha,\beta=1}^2 \sum_{I,J=1}^{6}
        x_{W{\tilde \chi}_\alpha^-}
        (\Gamma^d_{CL})^{\alpha 3}_I
        (\Gamma^{d\;\dagger}_{CL})_{\beta 2}^I
\\ \nonumber
  &&~~~~~~~~~(\Gamma^\nu_{CL})^{\alpha i}_J
        (\Gamma^{\nu\;\dagger}_{CL})_{\beta  i}^J
        2\sqrt{x_{{\tilde \chi}_\beta^- {\tilde \chi}_\alpha^-}}
        g_5(x_{{\tilde u}_I      {\tilde \chi}_\alpha^-},
           x_{{\tilde \ell}_J    {\tilde \chi}_\alpha^-},
           x_{{\tilde \chi}_\beta^- {\tilde \chi}_\alpha^-}),
\\
  C_{11,box}^{{\tilde \chi}^0}
  &=&\frac{1}{4}
    \sum_{\alpha,\beta=1}^{4} \sum_{I=1}^{6} \sum_{J=1}^{3}
        x_{W{\tilde \chi}_\alpha^0}
        (\Gamma^d_{NL})^{\alpha 3}_I
        (\Gamma^{d\;\dagger}_{NL})_{\beta 2}^I
\\\nonumber&&~~~~~
    \times \left[
       - (\Gamma^\nu_{NL})^{\alpha  i}_J
         (\Gamma^{\nu\;\dagger}_{NL})_{\beta i}^J
      2 \sqrt{x_{{\tilde \chi}_\beta^0 {\tilde \chi}_\alpha^0}}
      g_5(x_{{\tilde \nu}_J{\tilde \chi}^0_\alpha}
          ,x_{{\tilde d}_I{\tilde \chi}^0_\alpha}
          ,x_{{\tilde \chi}^0_\alpha {\tilde \chi}^0_\alpha})
    \right.
\\ \nonumber&&~~~~~
    +\left.
        (\Gamma^\nu_{NL})^{\beta i}_J
        (\Gamma^{\nu\;\dagger}_{NL})_{\alpha  i}^J
      g_6(x_{{\tilde \nu}_J{\tilde \chi}^0_\alpha}
          ,x_{{\tilde d}_I{\tilde \chi}^0_\alpha}
          ,x_{{\tilde \chi}^0_\beta {\tilde \chi}^0_\alpha})
    \right].
\end{eqnarray}

\section{Feynman rules}
\label{sec:feynman}
In this appendix, we give our notations.
The mass matrix of chargino is given by

\begin{eqnarray}
  \label{chargino}
  {\cal L} = -\left({\tilde W}^- ~ {\tilde h}_1^-\right)
             \left( 
               \begin{array}{cc}
                 M_2 & \sqrt{2} m_W \sin\beta \\
                 \sqrt{2} m_W \cos\beta & \mu \\
               \end{array}
             \right)
             \left(
               \begin{array}{c}
                 {\tilde W}^+  \\
                 {\tilde h}^+_2
               \end{array}
             \right) + {\rm H.c}.
\end{eqnarray}
Diagonalizing this mass matrix, mass eigenstates of chargino 
$\chi_i (i=1,2)$ are given by

\begin{eqnarray}
  \label{chmass}
  \left(
    \begin{array}{c}
      {\tilde \chi}_1^+  \\
      {\tilde \chi}_2^+  \\
    \end{array}
  \right)
  =
  U_+
  \left(
    \begin{array}{c}
      {\tilde W}^+    \\
      {\tilde h}^+_2  \\
    \end{array}
  \right)
~,~~~~
  \left(
    \begin{array}{c}
      {\tilde \chi}_1^-  \\
      {\tilde \chi}_2^-  \\
    \end{array}
  \right)
  =
  U_-
  \left(
    \begin{array}{c}
      {\tilde W}^-    \\
      {\tilde h}^-_1  \\
    \end{array}
  \right).
\end{eqnarray}

The mass matrix of neutralino is given by

\begin{eqnarray}
  \label{neutralino}
  {\cal L} = -\frac{1}{2}\left(
                {\tilde B} ~ {\tilde W}^3 ~
                {\tilde h}_1^0 ~ {\tilde h}_2^0
             \right)
             \left(
               \begin{array}{cccc}
                 M_1 & 0 & -m_Z\,s_W\,c_\beta & m_Z\,s_W\,s_\beta\\
                 0 & M_2 & m_Z\,c_W\,c_\beta & -m_Z\,c_W\,s_\beta\\
                 -m_Z\,s_W\,c_\beta & m_Z\,c_W\,c_\beta & 0 & -\mu \\
                 m_Z\,s_W\,s_\beta & -m_Z\,c_W\,s_\beta & -\mu & 0\\
               \end{array}
             \right)
             \left(
               \begin{array}{c}
                 {\tilde B}   \\
                 {\tilde W}^3 \\
                 {\tilde h}_1^0 \\
                 {\tilde h}_2^0 \\
               \end{array}
             \right),~~~
\end{eqnarray}
where $s_W = \sin\theta_W, c_W = \cos\theta_W, s_\beta = \sin\beta$ and 
$c_\beta = \cos\beta$.
Diagonalizing this mass matrix, mass eigenstates of neutralino are given
by

\begin{eqnarray}
  \label{numass}
  \left(
    \begin{array}{c}
      {\tilde \chi}_0^1 \\
      {\tilde \chi}_0^2 \\
      {\tilde \chi}_0^3 \\
      {\tilde \chi}_0^4 \\
    \end{array}
  \right)
  =
  U_N
  \left(
    \begin{array}{c}
      {\tilde B}   \\
      {\tilde W}^3 \\
      {\tilde h}_1^0 \\
      {\tilde h}_2^0 \\
    \end{array}
    \right).
\end{eqnarray}

The relevant interaction Lagrangian for the \bsl\,
process is written as follows.

\begin{itemize}
\item chargino-quark(lepton)-squark(slepton) interaction
  \begin{eqnarray}
    \label{cqs}
    {\cal L} &=&
    - g \, \overline{{\tilde \chi}^-_\alpha}
    \left[
      (\Gamma^d_{CL})^{\alpha j}_IP_L + (\Gamma^d_{CR})^{\alpha j}_IP_R
    \right]
    d_j {\tilde u}_I^\ast
\\ \nonumber
    &&
    - g \, \overline{{\tilde \chi}^-_\alpha}
    \left[
      (\Gamma^\ell_{CL})^{\alpha j}_IP_L + (\Gamma^\ell_{CR})^{\alpha j}_IP_R
    \right]
    \ell_j {\tilde \nu}_I^\ast 
\\ \nonumber
    &&
    - g \, \overline{({\tilde \chi}^-)^C_\alpha}
      (\Gamma^\nu_{CL})^{\alpha j}_IP_L
    \nu_j {\tilde \ell}_I^\ast + {\rm H.c.},
\\ \nonumber 
    P_{R(L)} &=& \frac{1}{2}(1\pm\gamma_5).
  \end{eqnarray}
  The  mixing matrices $\Gamma^d_{CL(R)}$, $\Gamma^\ell_{CL(R)}$ and 
$\Gamma^\nu_{CL}$ are given by
  \begin{eqnarray}
    \label{chmixing}
    (\Gamma^d_{CL})^{\alpha j}_I &=&
      ({\tilde U}_U)_I^j (U_+)^{\ast\alpha}_1 - ({\tilde U}_U)_I^{k+3}
      \frac{m_k^{(u)}}{\sqrt{2}m_W\sin\beta}(U_+)^{\ast\alpha}_2
    V_{kj},
\\
    (\Gamma^d_{CR})^{\alpha j}_I &=& 
    ({\tilde U}_U)_I^{k}
    \frac{m_k^{(d)}}{\sqrt{2}m_W\cos\beta}(U_-)_2^\alpha,
\\
    (\Gamma^\ell_{CL})^{\alpha j}_I &=& 
     ({\tilde U}_\nu)_I^j (U_+)_1^{\ast\alpha},
\\
    (\Gamma^\nu_{CL})^{\alpha j}_I &=& 
     ({\tilde U}_\ell)_I^j (U_-)_1^\alpha,
  \end{eqnarray}
where the matrices ${\tilde U}_U$, ${\tilde U}_\ell$ and ${\tilde U}_\nu$ are 
the unitary matrices which diagonalize
the up-type squark mass matrix, the slepton mass matrix and
 the sneutrino mass matrix respectively. $V$ is the CKM matrix.
Note that we neglect small contribution proportional to
the Yukawa couplings of the lepton.

\item neutralino-quark(lepton)-squark(slepton) interaction
  \begin{eqnarray}
    \label{nqs}
    {\cal L} &=&
    -g \, \overline{{\tilde \chi}^0_\alpha}
    \left[
      (\Gamma^d_{NL})^{\alpha j}_IP_L + (\Gamma^d_{NR})^{\alpha j}_IP_R
    \right]
    d_j {\tilde d}_I^\ast
\\ \nonumber
    &&-
    g \, \overline{{\tilde \chi}^0_\alpha}
    \left[
      (\Gamma^\ell_{NL})^{\alpha j}_IP_L + (\Gamma^\ell_{NR})^{\alpha j}_IP_R
    \right]
    \ell_j {\tilde \ell}_I^\ast
\\ \nonumber
    &&-
    g \, \overline{{\tilde \chi}^0_\alpha}
      (\Gamma^\nu_{NL})^{\alpha j}_IP_L
    \nu_j {\tilde \nu}_I^\ast + {\rm H.c.}
  \end{eqnarray}
The mixing matrices $\Gamma^d_{NL(R)}$, $\Gamma^\ell_{NL(R)}$ and 
$\Gamma^\nu_{NL}$ are given by 
\begin{eqnarray}
  \label{numxing}
  (\Gamma_{NL}^{d})_I^{\alpha j}
  &=& \sqrt{2}
    \left[
      \frac{1}{2}(U_N)_2^\alpha - \frac{1}{6}\tan\theta_W (U_N)_1^\alpha
    \right]
    ({\tilde U}_D)^j_I 
\\ \nonumber~~~~
   && -\frac{m^{d}_j}{\sqrt{2}m_W\cos\beta}
     (U_N)_3^\alpha({\tilde U}_D)_I^{j+3},
\\    
  (\Gamma_{NR}^{d})_I^{\alpha j}
  &=& \sqrt{2}
    \left[
      - \frac{1}{3}\tan\theta_W (U_N^\dagger)_1^\alpha
    \right]
    ({\tilde U}_D)_I^{j+3}
\\ \nonumber ~~~~
    &&-\frac{m^{d}_j}{\sqrt{2}m_W\cos\beta}
     (U_N^\dagger)_3^\alpha({\tilde U}_D)_I^{j},
\\
  (\Gamma_{NL}^{\ell})_I^{\alpha j}
  &=& \sqrt{2}
    \left[
      \frac{1}{2}(U_N)_2^\alpha + \frac{1}{2}\tan\theta_W (U_N)_1^\alpha
    \right]
    ({\tilde U}_L^\dagger)_I^j,
\\    
  (\Gamma_{NR}^{\ell})_I^{\alpha j}
  &=& \sqrt{2}
    \left[
      - \tan\theta_W (U_N^\dagger)_1^\alpha
    \right]
    ({\tilde U}_\ell^\dagger)_I^{j+3},
\\
  G_{\alpha\beta} &=& (U_N^\dagger)_\alpha^3(U_N)_3^\beta
                   -(U_N^\dagger)_\alpha^4(U_N)_4^\beta,
\\
  (\Gamma_{NL}^{(\nu)})_I^{\alpha j}
  &=& \sqrt{2}
    \left[
      -\frac{1}{2}(U_N)_2^\alpha + \frac{1}{2}\tan\theta_W (U_N)_1^\alpha
    \right]
    ({\tilde U}_\nu^\dagger)_I^j,
\end{eqnarray}
where the matrix ${\tilde U}_D$
is the unitary matrix which diagonalizes
the down-type squark mass matrix.

\item gluino-quark-squark interaction
  \begin{eqnarray}
    \label{gqs}
    {\cal L} &=&
    -g_s \sqrt{2}\, (T^a)_{\alpha\beta}{\overline {\tilde G^a}}
    \left[
      (\Gamma^d_{GL})^{j}_IP_L + (\Gamma^d_{GR})^{j}_IP_R
    \right]
    d_{j\alpha} {\tilde d}_{I\beta}^\ast,
  \end{eqnarray}
where the mixing matrices $\Gamma^d_{GL(R)}$ are given by
\begin{eqnarray}
  \label{glmixing}
  (\Gamma^d_{GL})^j_I &=& ({\tilde U}_D)_I^j,
\\
  (\Gamma^d_{GR})^j_I &=& -({\tilde U}_D)_I^{j+3}.
\end{eqnarray}
\end{itemize}

\section{One-loop functions}
\label{sec:function}
These are the one loop functions which appear in calculating
the penguin or box diagrams.
\begin{eqnarray}
  \label{function}
  f_1(x)   &=&
        \frac{1}{12(x-1)^4}(x^3-6x^2+3x+2+6x\ln x),
\\
  f_2(x)   &=&
        \frac{1}{12(x-1)^4}(2x^3+3x^2-6x+1-6x^2\ln x),
\\
  f_3(x)   &=&
        \frac{1}{2(x-1)^3}(x^2-4x+3+2\ln x),
\\
  f_4(x)   &=&
        \frac{1}{2(x-1)^3}(x^2-1+2x\ln x),
\\
  f_5(x)   &=&
        \frac{1}{36(x-1)^4}(7x^3-36x^2+45x-16+(18x-12)\ln x),
\\
  f_6(x)   &=&
        \frac{1}{36(x-1)^4}(-11x^3+18x^2-9x+2+6x^3\ln x),
\\
  f_7(x)   &=&
        \frac{1}{12(x-1)^4}(x^3+10x^2-29x+18-(8x^2-6x-8)\ln x),
\\
  f_8(x)   &=&
        \frac{1}{12(x-1)^4}(-7x^3+8x^2+11x-12-(2x^3-20x^2+24x)\ln x),~~~~~~
\\
  f_9(x)   &=&
        \frac{1}{2(x-1)^2}(x^2-7x+6+(3x+2)\ln x),
\\
  g_1(x,y) &=&
        \frac{1}{x-y}
        \left[\frac{x}{x-1}\ln x - (x \leftrightarrow y)\right],
\\
  g_2(x,y) &=&
        \frac{1}{x-y}
        \left[\frac{x^2}{x-1}\ln x -\frac{3}{2}x
        - (x \leftrightarrow y)\right],
\\
  g_3(x,y) &=&
        \frac{1}{x-y}
        \left[\frac{x^2}{(x-1)^2}\ln x-\frac{1}{x-1}
         - (x \leftrightarrow y)\right],
\\
  g_5(x,y,z) &=&
        -\frac{1}{x-y}
        \left[ \frac{1}{x-z} 
                \left[ \frac{x}{x-1} \ln x 
                 - (x  \leftrightarrow z) \right]
                -(x \leftrightarrow y) \right],
\\
  g_6(x,y,z) &=&
        \frac{1}{x-y}
        \left[ \frac{1}{x-z} 
                \left[ \frac{x^2}{x-1} \ln x  -\frac{3 x}{2}
                 - (x  \leftrightarrow z) \right]
                -(x \leftrightarrow y) \right].
\end{eqnarray}

\section{The QCD correction}
\label{sec:QCD}
With $C_i(m_W)$ as the initial condition, we obtain the solution of
RGE in the LLA approximation~\cite{Ali,LLA}.
\begin{eqnarray}
  \label{LLA}
C_1 (Q ) &=& \frac{1}{2} C_2 (m_W)
               \left( \eta^{6/23} -  \eta^{-12/23} \right),
\\
C_2 (Q ) &=& \frac{1}{2} C_2 (m_W)
               \left( \eta^{6/23} +  \eta^{-12/23} \right),
\\
C_3 (Q ) &=& C_2 (m_W) \left(-0.0112 \eta^{-0.8994}
                         + \frac{1}{6} \eta^{-12/23}
                         - 0.1403 \eta^{-0.4230}
                         + 0.0054 \eta^{0.1456} \right.
\\ \nonumber
           && \qquad \qquad  \left. \vphantom{\frac{1}{6}}
                         - 0.0714 \eta^{6/23}
                         + 0.0509 \eta^{0.4086} \right),
\\
C_4 (Q) &=& C_2 (m_W) \left(0.0156 \eta^{-0.8994}
                         - \frac{1}{6} \eta^{-12/23}
                         + 0.1214 \eta^{-0.4230}
                         + 0.0026 \eta^{0.1456} \right.
\\ \nonumber
           && \qquad \qquad  \left. \vphantom{\frac{1}{6}}
                         - 0.0714 \eta^{6/23}
                         + 0.0984 \eta^{0.4086} \right),
\\
C_5 (Q) &=& C_2 (m_W) \left(-0.0025 \eta^{-0.8994}
                         + 0.0117 \eta^{-0.4230}
                         + 0.0304 \eta^{0.1456} 
                         - 0.0397 \eta^{0.4086} \right),~~~~~~
\\
C_6 (Q) &=& C_2 (m_W) \left(-0.0462 \eta^{-0.8994}
                         + 0.0239 \eta^{-0.4230}
                         - 0.0112 \eta^{0.1456}
                         + 0.0335 \eta^{0.4086} \right),~~~~~~
\\
C_7 (Q) &=& C_7 (m_W) \eta^{16/23} +
              C_8 (m_W) \frac{8}{3}
                        \left( \eta^{14/23} - \eta^{16/23} \right)
\\ \nonumber
          && + C_2 (m_W) \left(- 0.0185 \eta^{-0.8994}
                               - 0.0714 \eta^{-12/23}
                               - 0.0380 \eta^{-0.4230}
                               - 0.0057 \eta^{0.1456} \right.
\\ \nonumber
           && \qquad \qquad  \left.
                               - 0.4286 \eta^{6/23}
                               - 0.6494 \eta^{0.4086}
                               + 2.2996 \eta^{14/23}
                               - 1.0880 \eta^{16/23} \right),
\\
C_8 (Q) &=& C_8 (m_W) \eta^{14/23}
\\ \nonumber
          && + C_2 (m_W) \left(- 0.0571 \eta^{-0.8994}
                               + 0.0873 \eta^{-0.4230}
                               + 0.0209 \eta^{0.1456} \right.
\\ \nonumber
           && \qquad \qquad  \left.
                               - 0.9135 \eta^{0.4086}
                               + 0.8623 \eta^{14/23} \right),
\\
C_9 (Q) &=& C_9 (m_W) + \frac{\pi}{\alpha_s (m_W)}C_2 (m_W) \left( -0.1875
                         + 0.1648 \eta^{1-0.8994}
                         + 0.2424 \eta^{1-12/23}
                          \right.
\\ 
           && \qquad \qquad  \left.
                         + 0.1384 \eta^{1-0.4230}
                         - 0.0073 \eta^{1+0.1456}
                         - 0.3941 \eta^{1+6/23}
                         + 0.0433 \eta^{1+0.4086}\right),
\nonumber \\
C_{10} (Q ) &=& C_{10} (m_W) ,
\nonumber \\
C_{11} (Q ) &=& C_{11} (m_W) ,
\end{eqnarray}
where $\eta = \frac{\alpha_s (m_W)}{\alpha_s (Q)}$.

\section{The kinematical functions}
\label{sec:kfunc}

In this appendix we show the explicit form of kinematical functions
in Eq.~\ref{decayrate} and Eq.~\ref{AFB}.
 Here $\hat{s}$, $\hat{m}_s$ and $\hat{m}_\ell$ 
means $s/m_b^2$, $m_s/m_b$ and $m_\ell/m_b$.
\begin{eqnarray}
w(\hat{s})&=&\sqrt{(\hat{s}-(1 + \hat{m}_s )^2)(\hat{s}-(1-\hat{m}_s
    )^2)},
\\
\alpha_1 (\hat{s},\hat{m}_s,\hat{m}_\ell) 
&=& \left(1+\frac{2 \hat{m}_\ell^2}{\hat{s}}\right)
\left(- 2 \hat{s}^2 + \hat{s} (1+ \hat{m}_s^2)+(1-\hat{m}_s^2)^2 \right),
\\
\alpha_2 (\hat{s},\hat{m}_s,\hat{m}_\ell)
&=&\left(- 2 \hat{s}^2 + \hat{s} (1+ \hat{m}_s^2)+(1-\hat{m}_s^2)^2\right)
\\ \nonumber
&&+\frac{2\hat{m}_\ell^2}{\hat{s}}
\left(4\hat{s}^2-5(1+\hat{m}_s^2)\hat{s}+(1-\hat{m}_s^2)^2\right),
\\
\alpha_3 (\hat{s},\hat{m}_s,\hat{m}_\ell) 
&=& \left(1+\frac{2\hat{m}_\ell^2}{\hat{s}}\right)
\\ \nonumber
&&\times
\left(-(1+ \hat{m}_s^2) \hat{s}^2
       - (1+14 \hat{m}_s^2+\hat{m}_s^4) \hat{s}
       +2 (1+ \hat{m}_s^2)(1-\hat{m}_s^2)^2 
\right),
\\
\alpha_4  (\hat{s},\hat{m}_s,\hat{m}_\ell) 
&=& \left(1+\frac{2\hat{m}_\ell^2}{\hat{s}}\right)
\left( (1-\hat{m}_s^2)^2 - (1+ \hat{m}_s^2) \hat{s} \right).
\end{eqnarray}


\clearpage

\section*{Figure Captions}

\begin{figure}[htbp]
    \leavevmode
    \caption{Feynman diagrams in the SM. (a) \bsg, (b) $
      b \rightarrow s\,g$, (c) the penguin diagram for \bsll,
      (d) box diagram for \bsll.
      The SUSY contributions to these diagrams are obtained 
      by replacing the internal lines with SUSY particles.}
    \label{fig:SM}
\end{figure}

\begin{figure}[htbp]
    \leavevmode
    \caption{ \cs, \cn\, and \ct\, in the SUGRA model normalized to that of 
      in the SM (a) for $\tan\beta\, = 3$ and (b) for $\tan\beta\, = 30$.}
    \label{fig:c7c9}
\end{figure}

\begin{figure}[htbp]
  \begin{center}
    \leavevmode
    \caption{\br(\bsll) in the SM and in the minimal SUGRA model for
      $\kappa=\pm 1$. The SUSY parameters are fixed with 
      $\tan\beta\, = 30$, $m_0 = 369\ {\rm GeV},$ $ M_{gX} = 100\ {\rm
        GeV},$ $A_X=m_0$, where \cs\ becomes the opposite sign to the SM.}
    \label{fig:spectrum}
  \end{center}
\end{figure}

\begin{figure}[htbp]
  \begin{center}
    \leavevmode
    \caption{ ${\cal A}_{\rm FB}$(\bsll)\,
      in the SM and in the minimal SUGRA model for $\kappa=\pm 1$.
       The SUSY
      parameters are fixed with $\tan\beta\, = 30,$
      $m_0=369\ {\rm GeV},$  $M_{gX}=100\ {\rm GeV},$ $A_X = m_0$, 
      where \cs\, becomes
      the opposite sign to the SM. }
    \label{fig:FBA}
  \end{center}
\end{figure}

\begin{figure}[htbp]
  \begin{center}
    \leavevmode
    \caption{ A correlation between  \br(\bsg)\
      and \br(\bsll)\
      in the minimal SUGRA model
      (a) in the low $s$ region with $\kappa = +1$,
      (b) in the low $s$ region with $\kappa = -1$, 
      (c) in the high $s$ region with $\kappa = +1$ and 
      (d) in the high $s$ region with $\kappa = -1$  
      for $\tan\beta\, = 30$. 
      Two vertical dashed line
      represent the experimental \bsg\, constraint.
      Circles, squares and triangles represent how much 
      \br( \bsg)\, and \br (\bsll)\,
      change when the renormalization point $Q$ is taken to be $
      m_b/2, m_b$ and $2 m_b$ respectively.}
    \label{fig:br30}
  \end{center}
\end{figure}

\begin{figure}[htbp]
  \begin{center}
    \leavevmode
    \caption{A correlation between \br(\bsg)\
      and ${\cal A}_{\rm FB}$( \bsll)\,
      in the minimal SUGRA model
      (a) in the low $s$ region with $\kappa = +1$,
      (b) in the low $s$ region with $\kappa = -1$, 
      (c) in the high $s$ region with $\kappa = +1$ and 
      (d) in the high $s$ region with $\kappa = -1$ 
      for $\tan\beta\, = 30$. 
      Two vertical dashed line
      represent the experimental \bsg\, constraint.
      Circles, squares and triangles represent how much 
      \br( \bsg)\ and
      ${\cal A}_{\rm FB}$(\bsll) \,
      change when the renormalization point $Q$ is taken to be $
      m_b/2, m_b$ and $2 m_b$ respectively.}
    \label{fig:bsfb}
  \end{center}
\end{figure}

\begin{figure}[htbp]
  \begin{center}
    \leavevmode
    \caption{ \br(\bsll)
      in the low $s$ region for \tb\,=\, 30 as a function
      of (a) the light chargino mass  and (b) the light stop mass.
      The solid line shows the value in the SM.}
    \label{fig:chbr}
  \end{center}
\end{figure}

\begin{figure}[htbp]
  \begin{center}
    \leavevmode
    \caption{ \br(\bsnn) 
      for \tb\,=\,30 as a function of (a) the light chargino mass 
      and (b) the light stop mass.
      The solid line shows the value in the SM.}
    \label{fig:bsnunu}
  \end{center}
\end{figure}

\clearpage

\pagestyle{empty}

\begin{figure}[p]
  \begin{center}
    \leavevmode
    \centerline{\psfig{file=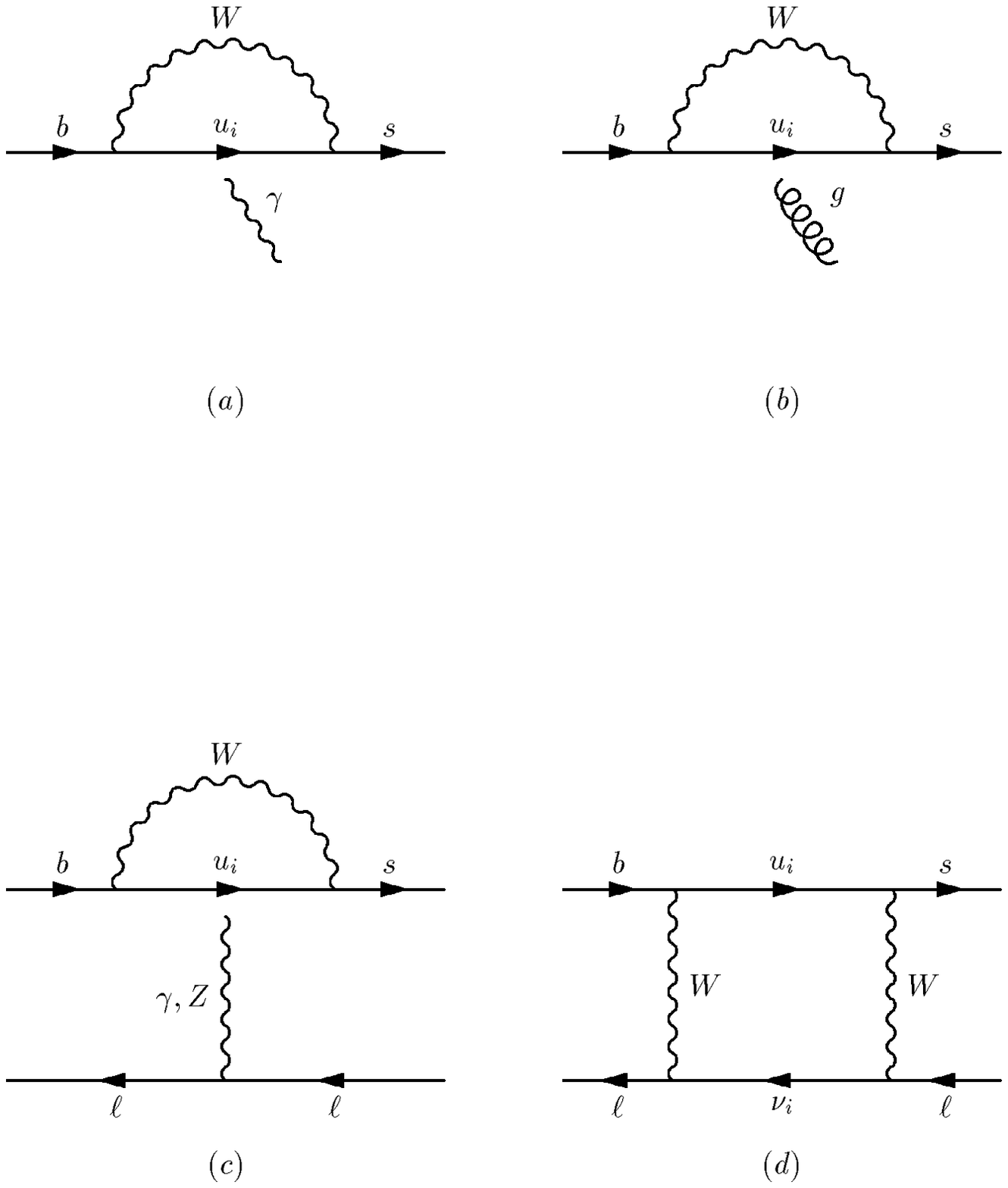,height=0.8\textheight}}
\\ 
    {\vskip 2cm}
    {\huge\bf Fig.~1}
  \end{center}
\end{figure}

\begin{figure}[p]
  \begin{center}
    \leavevmode
    \centerline{\psfig{file=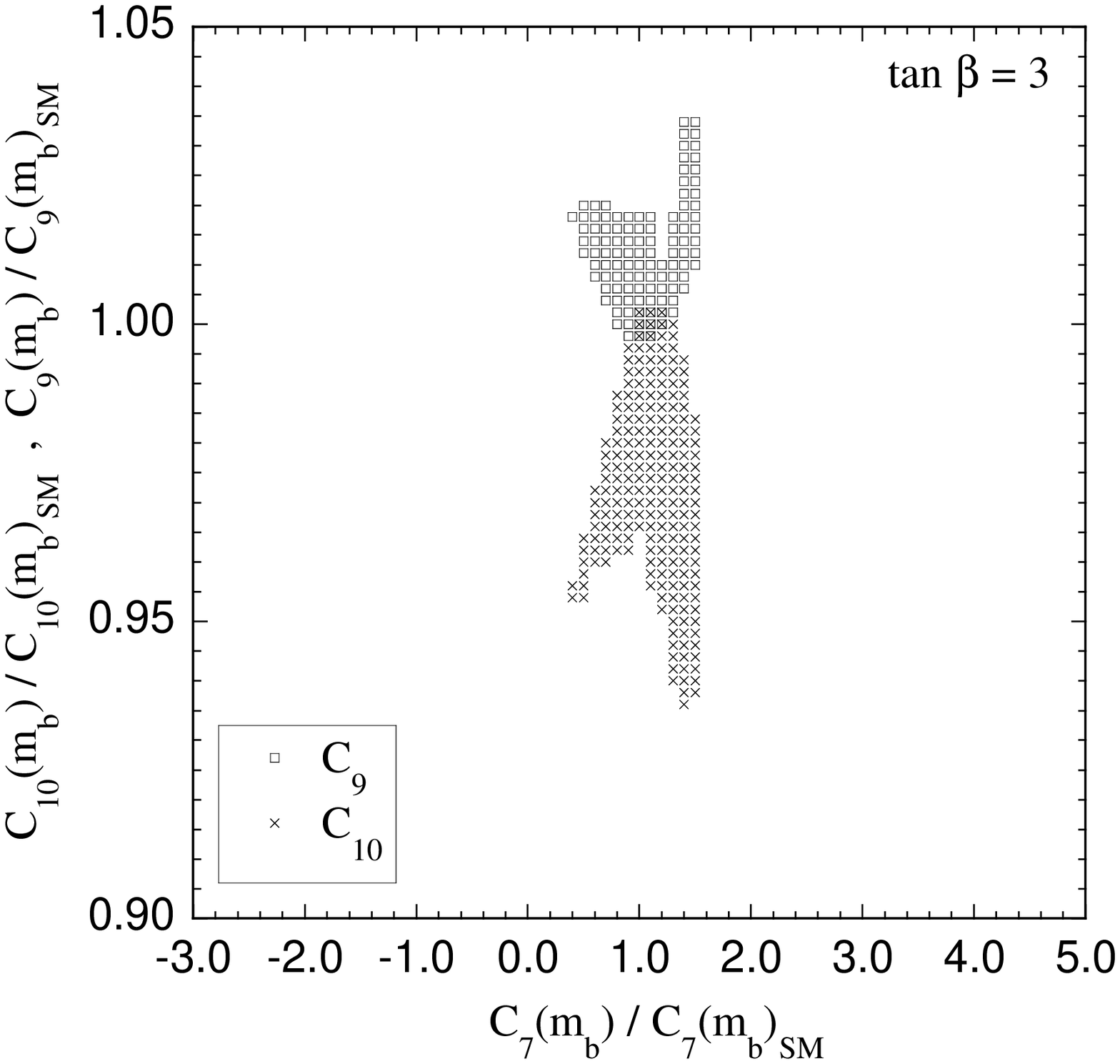}}
\\
    {\vskip 4cm}
    {\huge\bf Fig.~2 (a)}
  \end{center}
\end{figure}

\begin{figure}[p]
  \begin{center}
    \leavevmode
    \centerline{\psfig{file=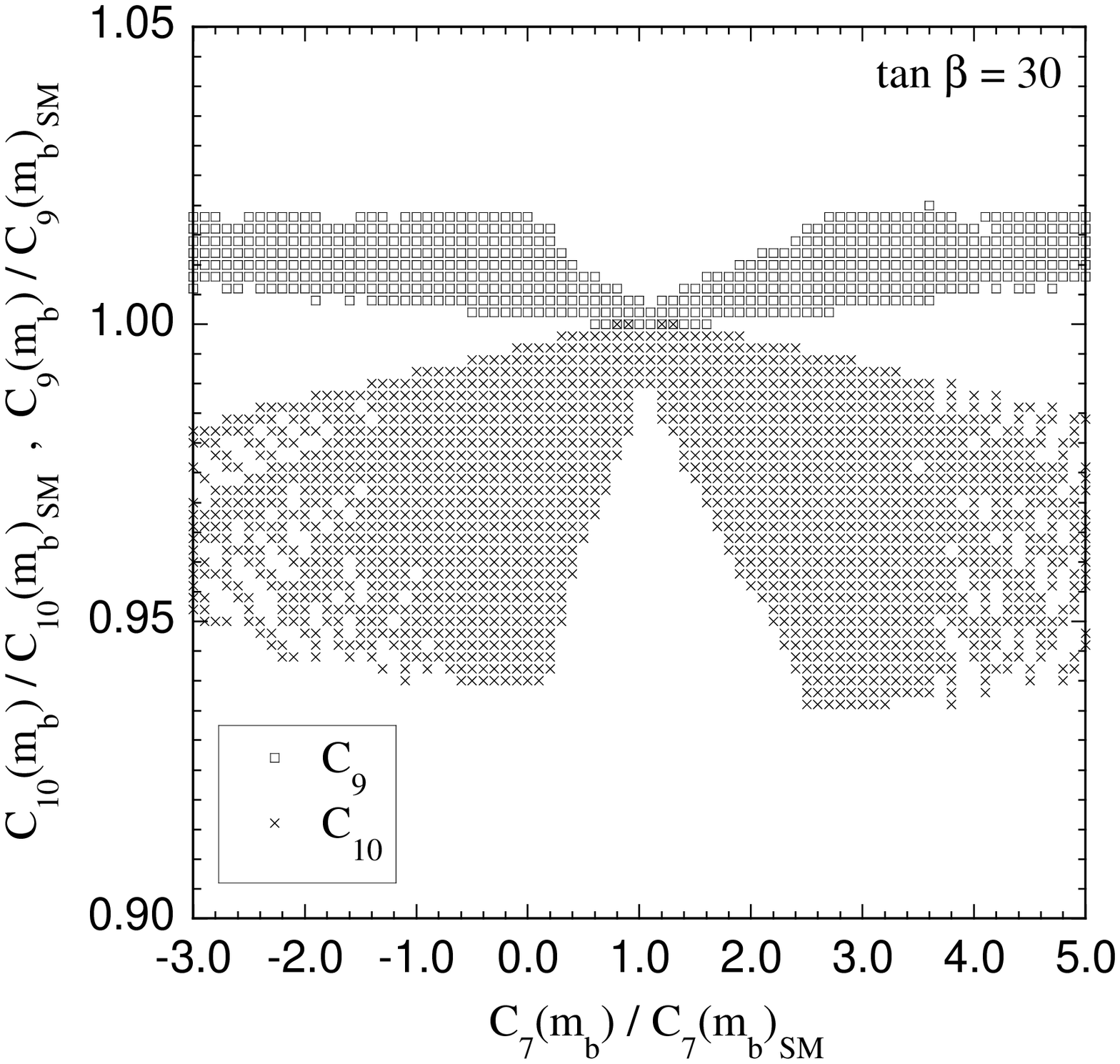}}
\\
    {\vskip 4cm}
    {\huge\bf Fig.~2 (b)}
  \end{center}
\end{figure}

\begin{figure}[p]
  \begin{center}
    \leavevmode
    \centerline{\psfig{file=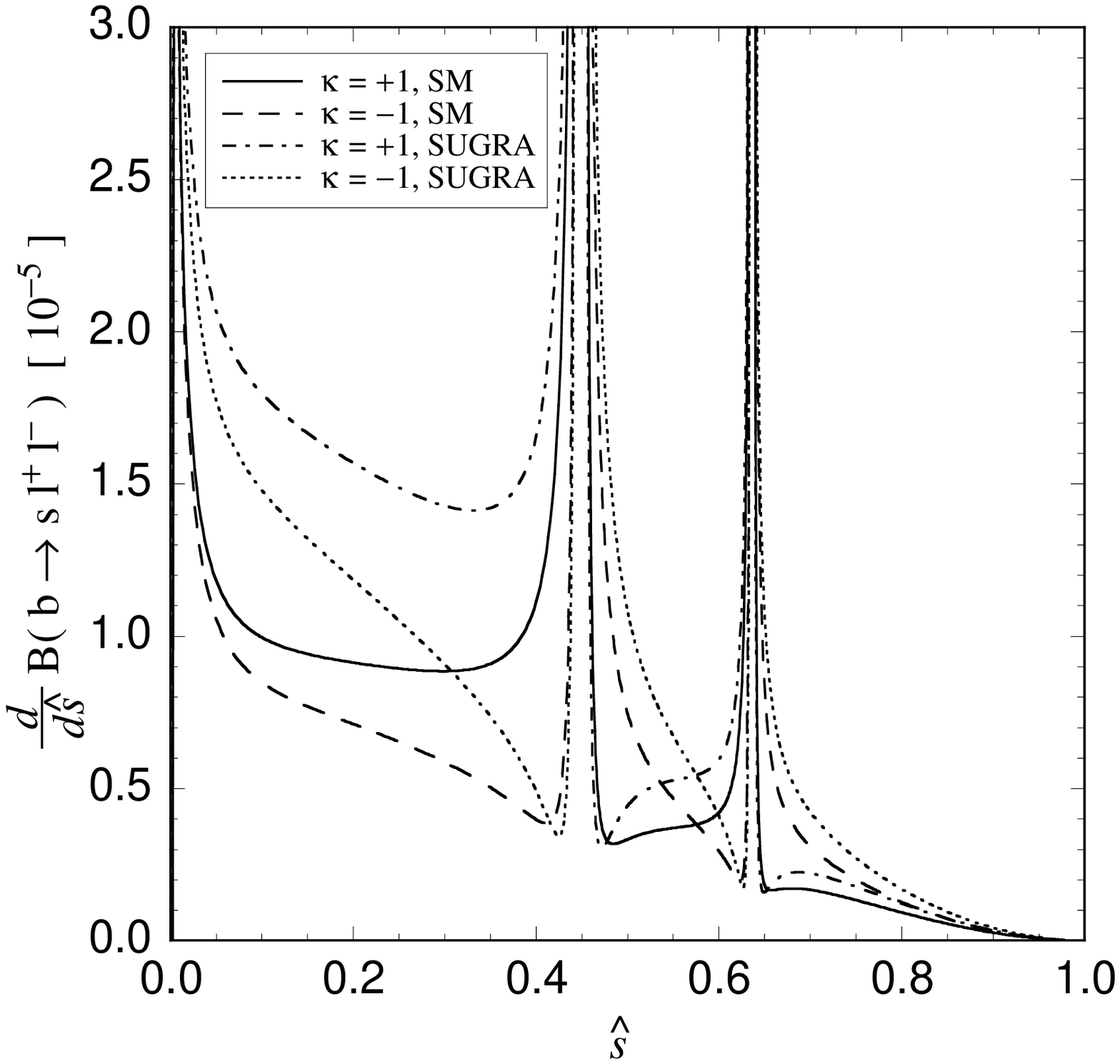}}
\\
    {\vskip 4cm}
    {\huge\bf Fig.~3}
  \end{center}
\end{figure}

\begin{figure}[p]
  \begin{center}
    \leavevmode
    \centerline{\psfig{file=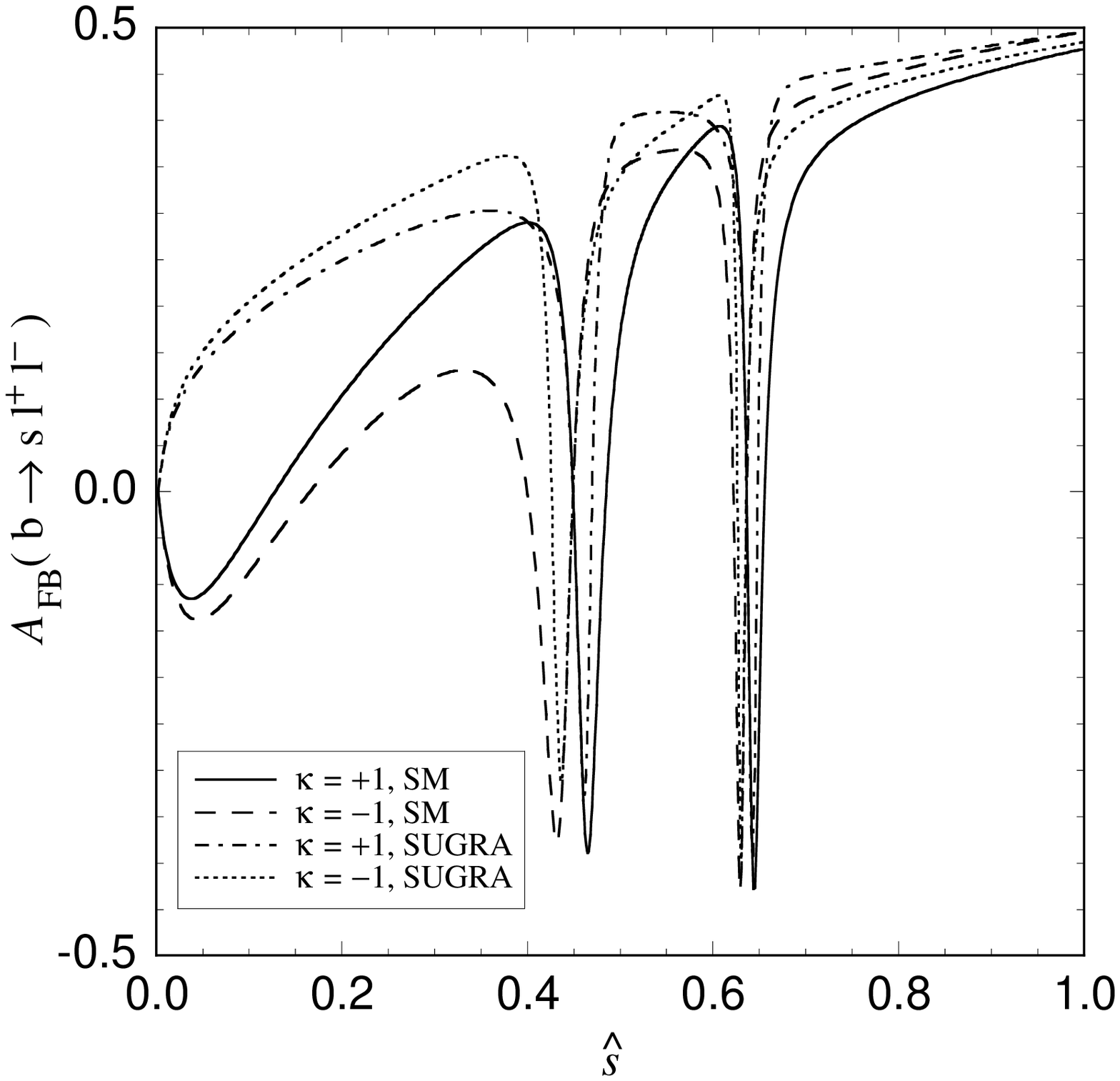}}
\\
    {\vskip 4cm}
    {\huge\bf Fig.~4}
  \end{center}
\end{figure}

\begin{figure}[p]
  \begin{center}
    \leavevmode
    \centerline{\psfig{file=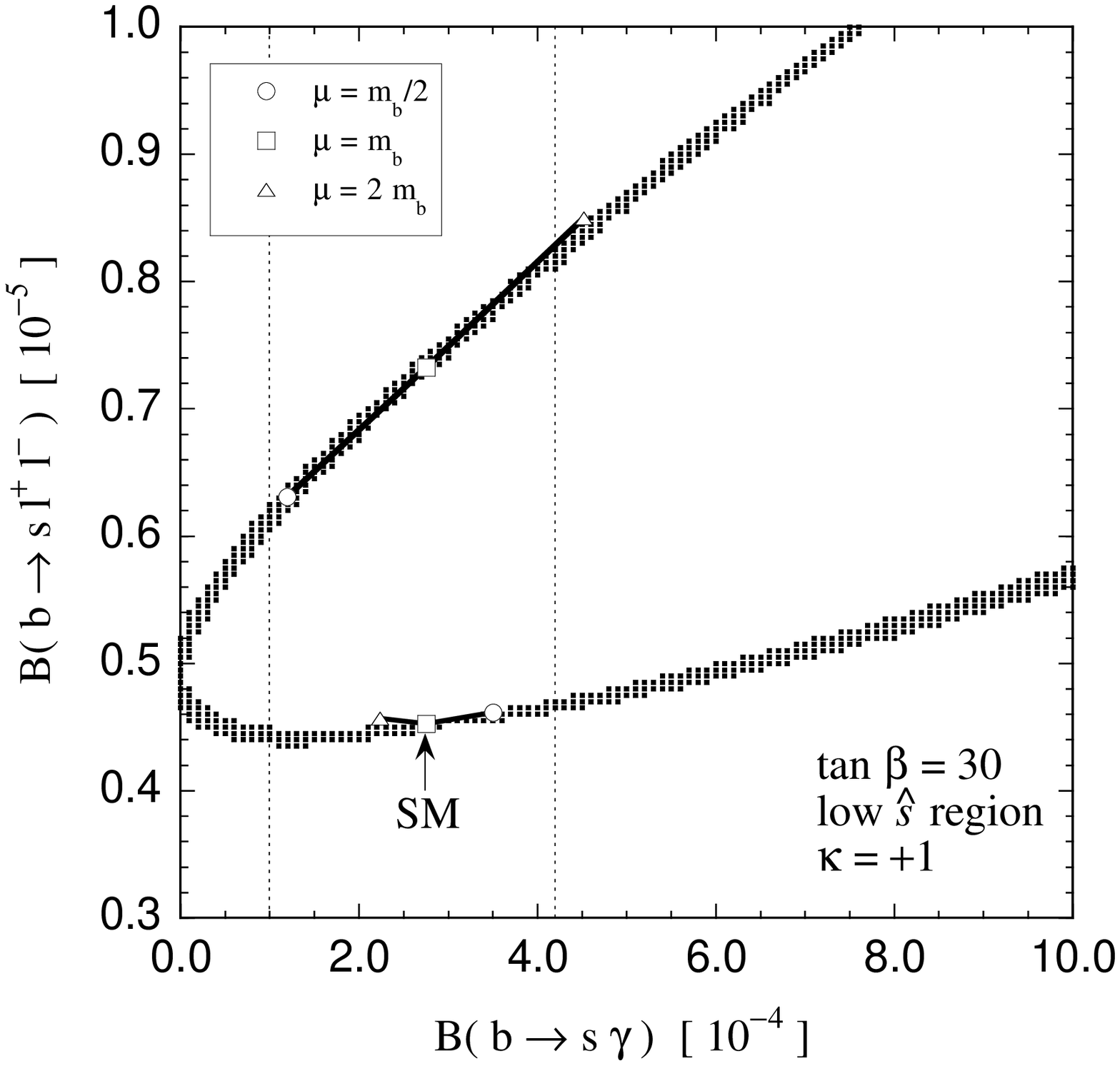}}
\\
    {\vskip 4cm}
    {\huge\bf Fig.~5 (a)}
  \end{center}
\end{figure}

\begin{figure}[p]
  \begin{center}
    \leavevmode
    \centerline{\psfig{file=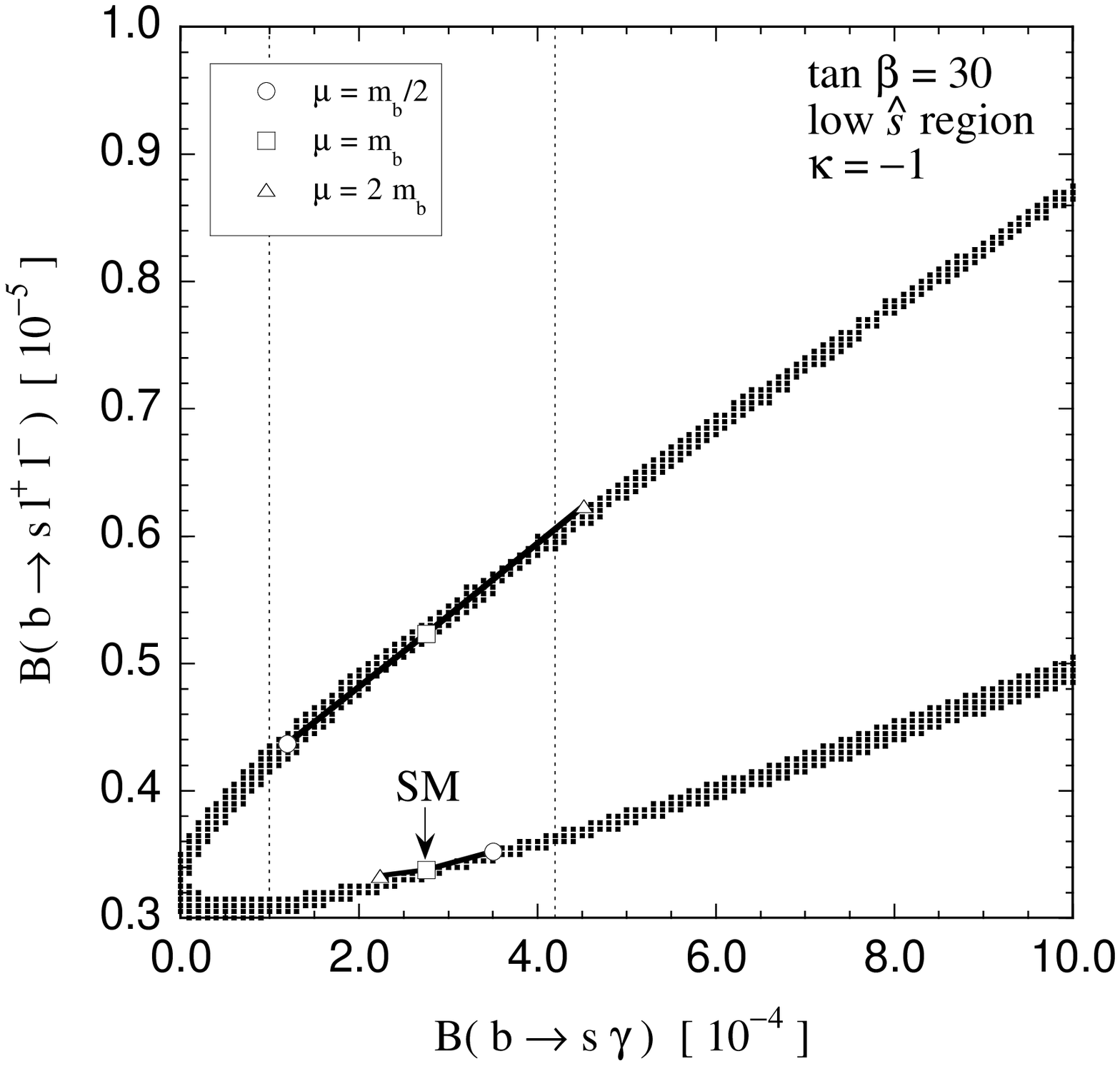}}    
\\
    {\vskip 4cm}
    {\huge\bf Fig.~5 (b)}
  \end{center}
\end{figure}

\begin{figure}[p]
  \begin{center}
    \leavevmode
    \centerline{\psfig{file=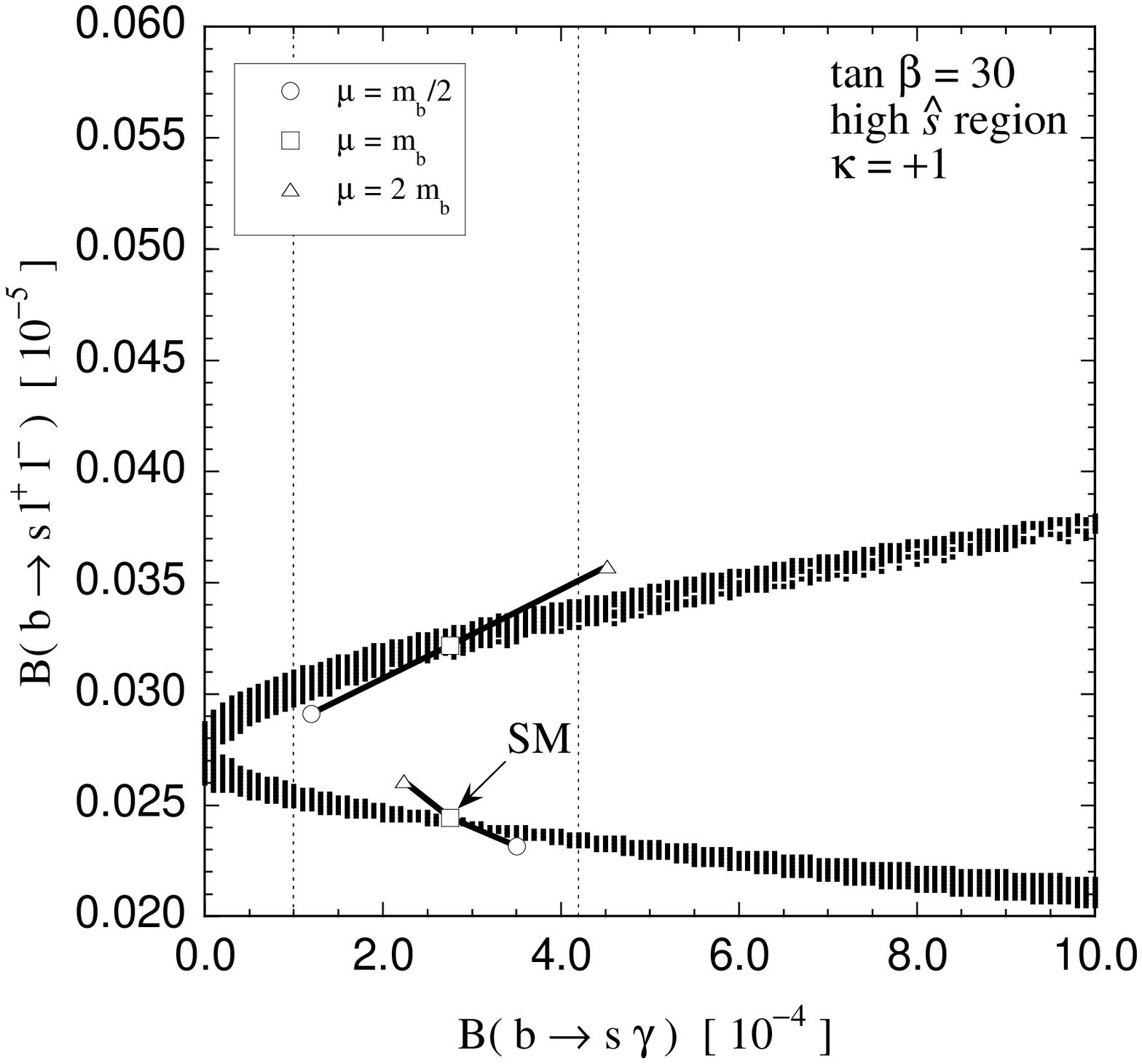}}
\\
    {\vskip 4cm}
    {\huge\bf Fig.~5 (c)}
  \end{center}
\end{figure}

\begin{figure}[p]
  \begin{center}
    \leavevmode
    \centerline{\psfig{file=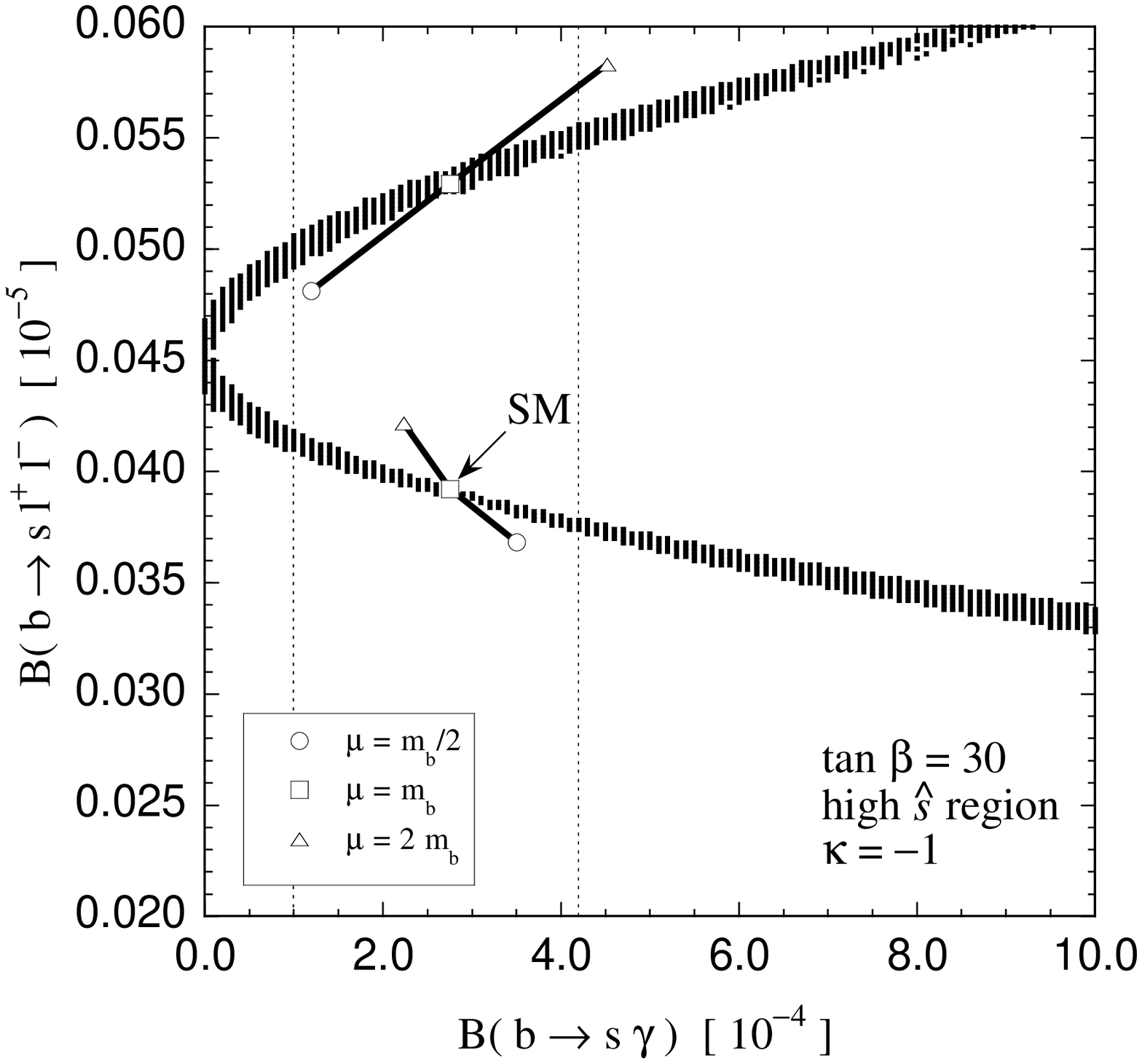}}
\\
    {\vskip 4cm}
    {\huge\bf Fig.~5 (d)}
  \end{center}
\end{figure}

\begin{figure}[p]
  \begin{center}
    \leavevmode
    \centerline{\psfig{file=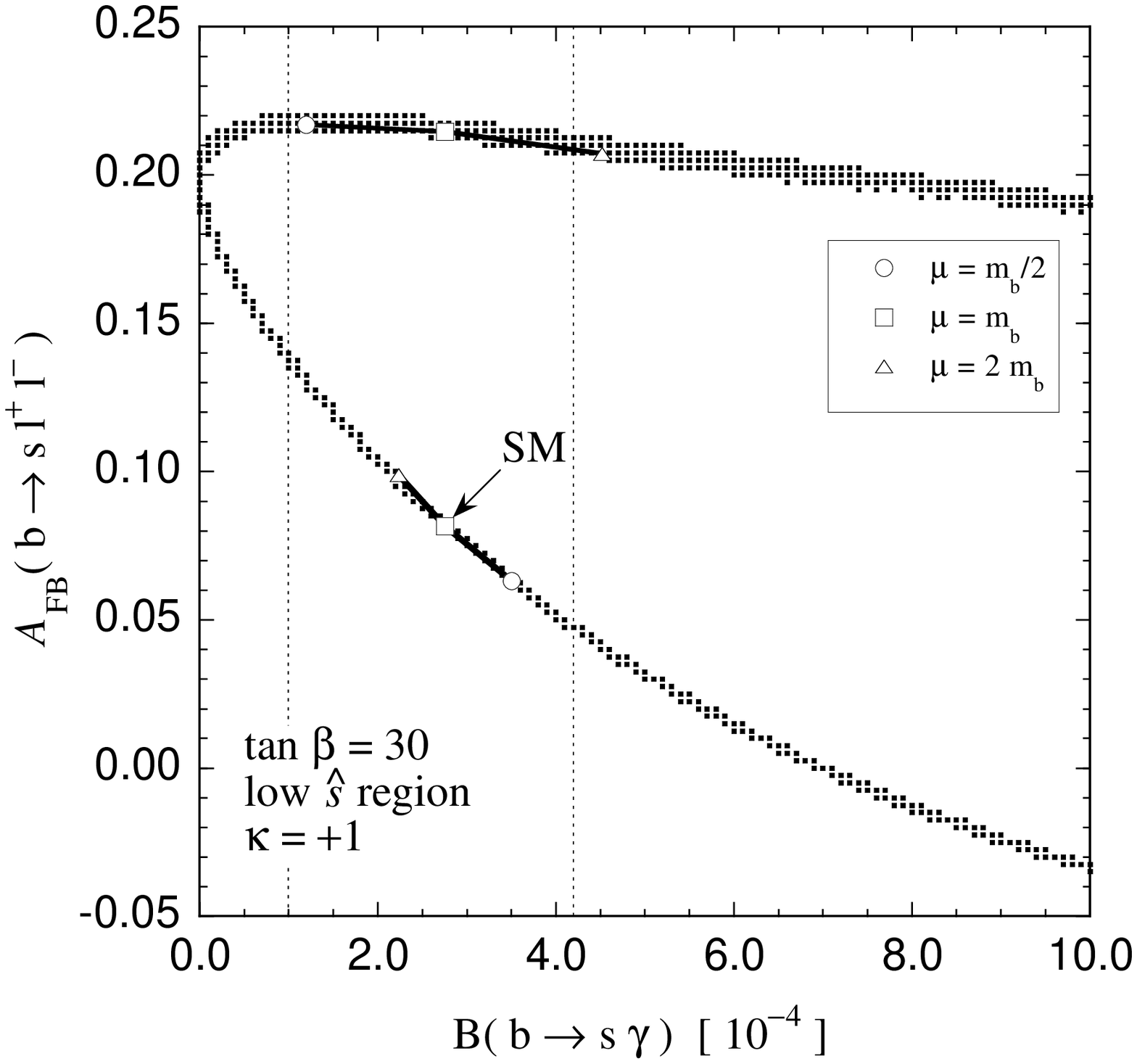}}
\\
    {\vskip 4cm}
    {\huge\bf Fig.~6 (a)}
  \end{center}
\end{figure}

\begin{figure}[p]
  \begin{center}
    \leavevmode
    \centerline{\psfig{file=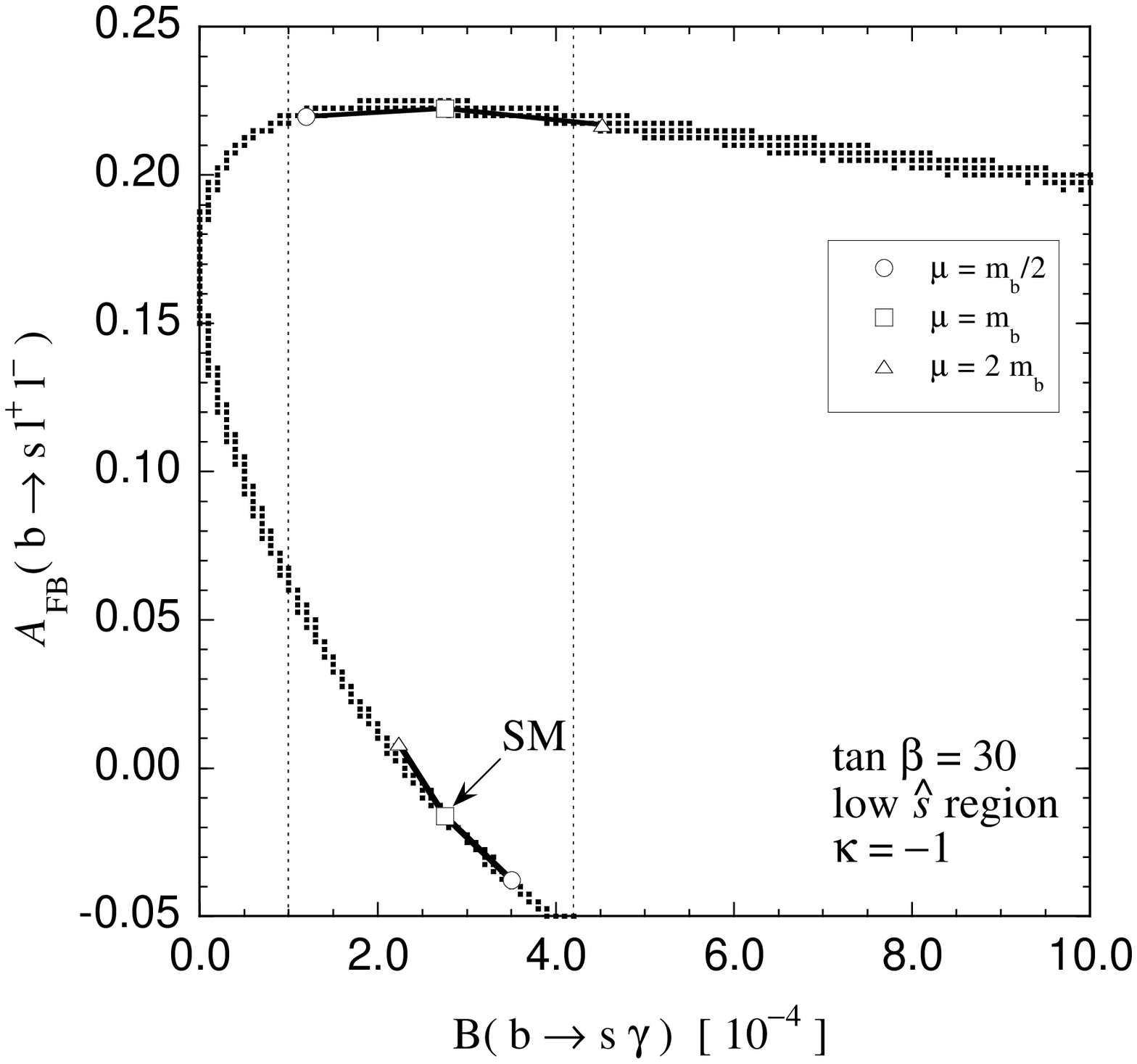}}    
\\
    {\vskip 4cm}
    {\huge\bf Fig.~6 (b)}
  \end{center}
\end{figure}

\begin{figure}[p]
  \begin{center}
    \leavevmode
    \centerline{\psfig{file=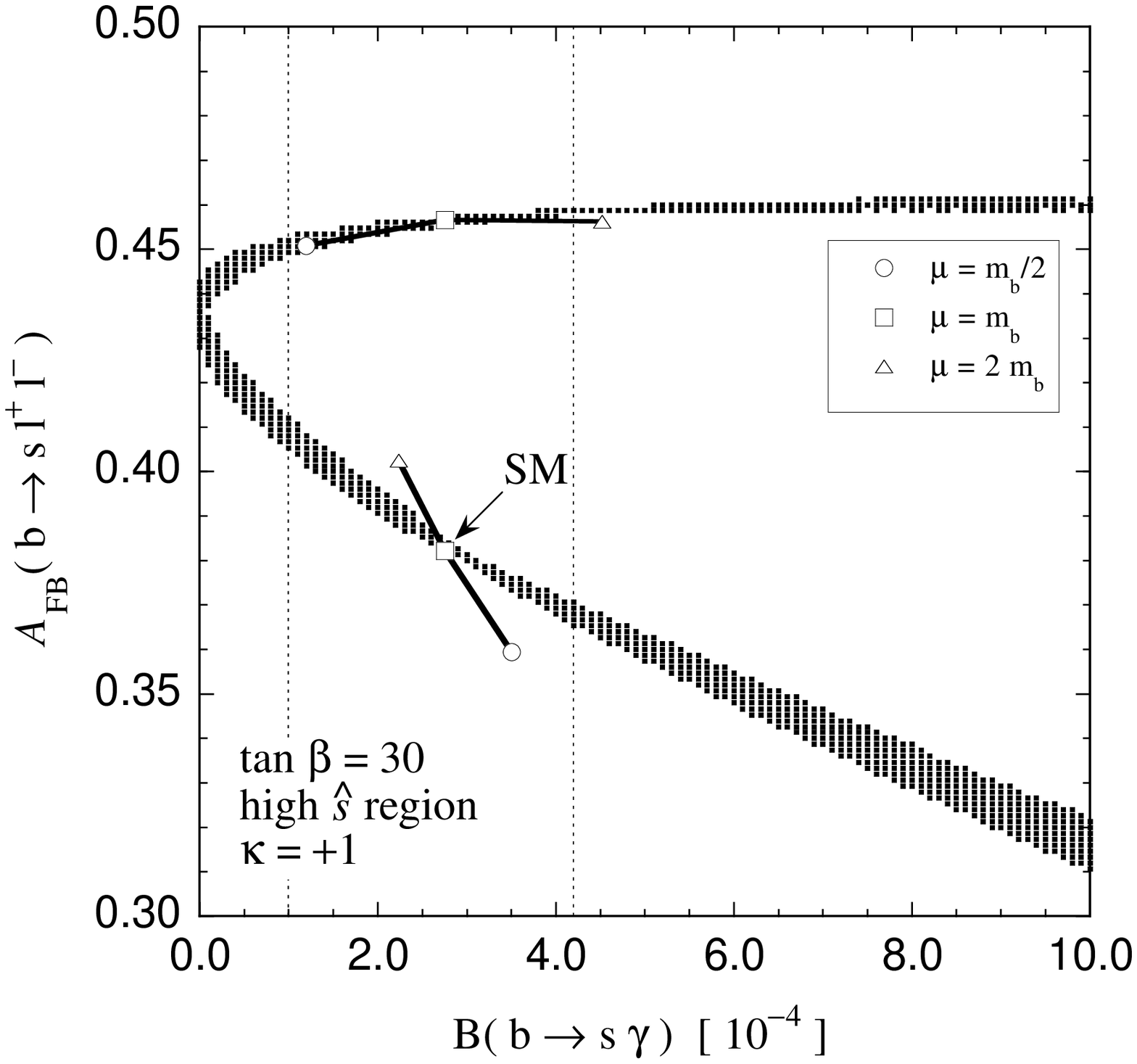}}
\\
    {\vskip 4cm}
    {\huge\bf Fig.~6 (c)}
  \end{center}
\end{figure}

\begin{figure}[p]
  \begin{center}
    \leavevmode
    \centerline{\psfig{file=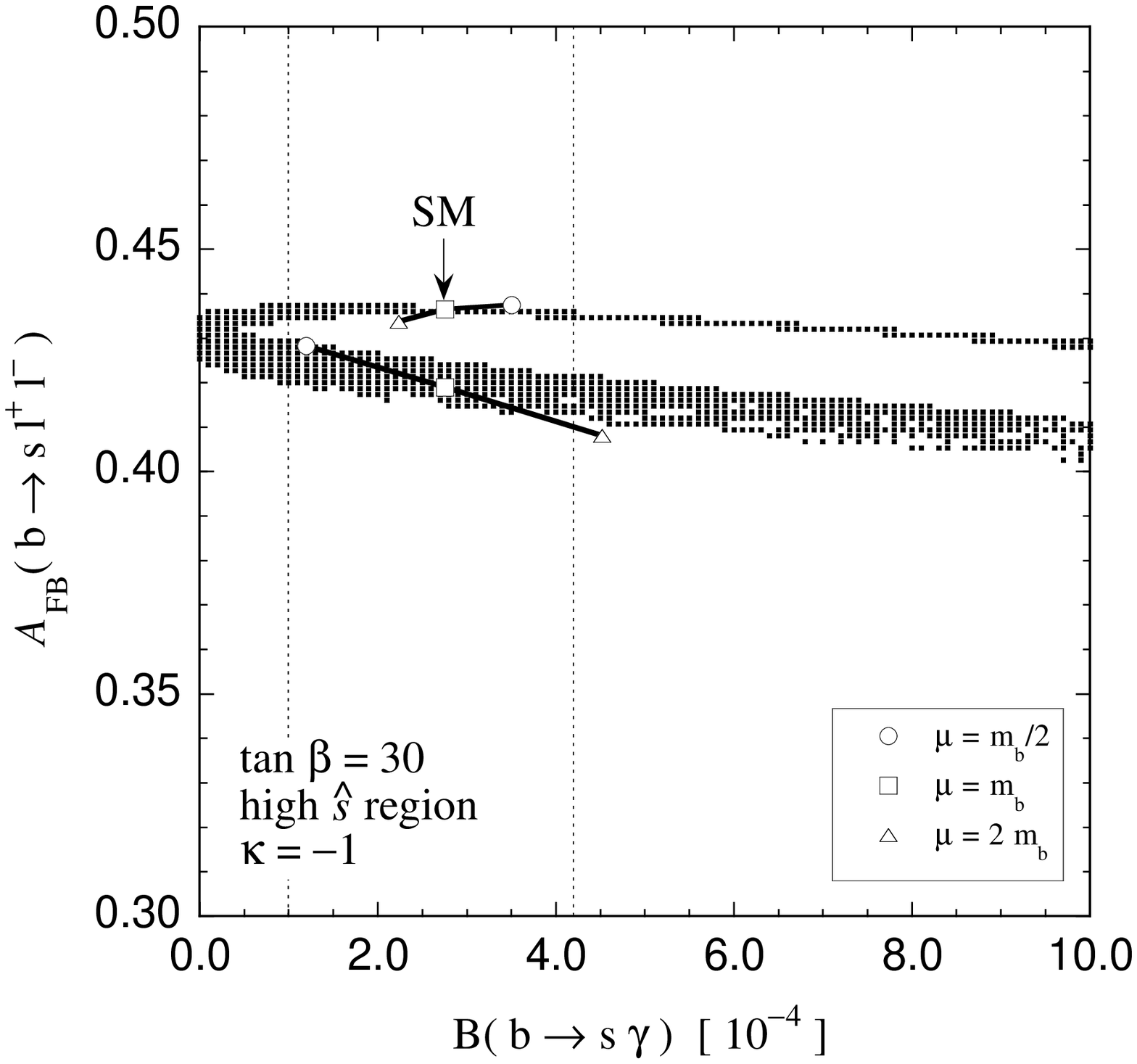}}
\\
    {\vskip 4cm}
    {\huge\bf Fig.~6 (d)}
  \end{center}
\end{figure}

\begin{figure}[p]
  \begin{center}
    \leavevmode
    \centerline{\psfig{file=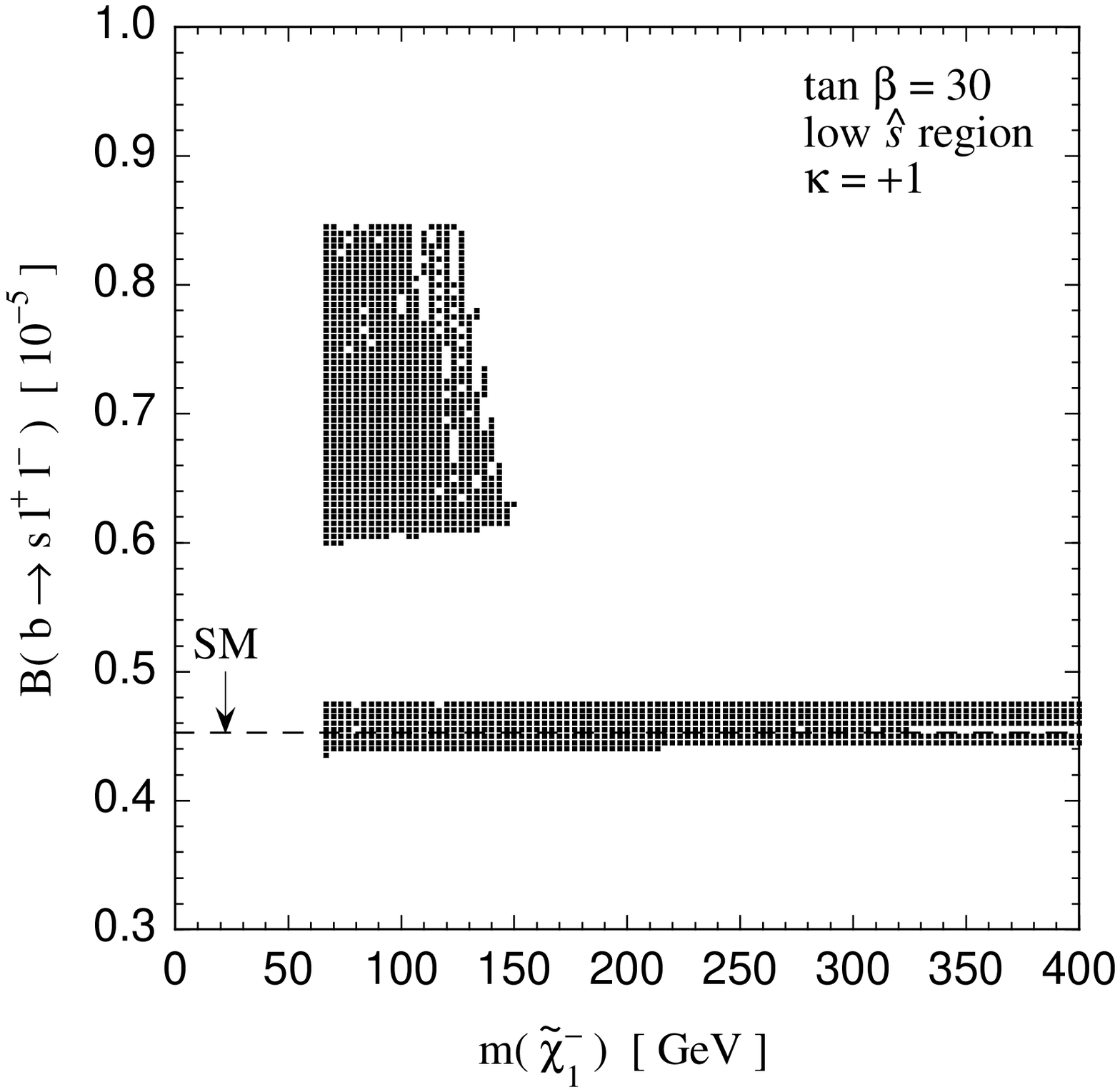}}
\\
    {\vskip 4cm}
    {\huge\bf Fig.~7 (a)}
  \end{center}
\end{figure}

\begin{figure}[p]
  \begin{center}
    \leavevmode
    \centerline{\psfig{file=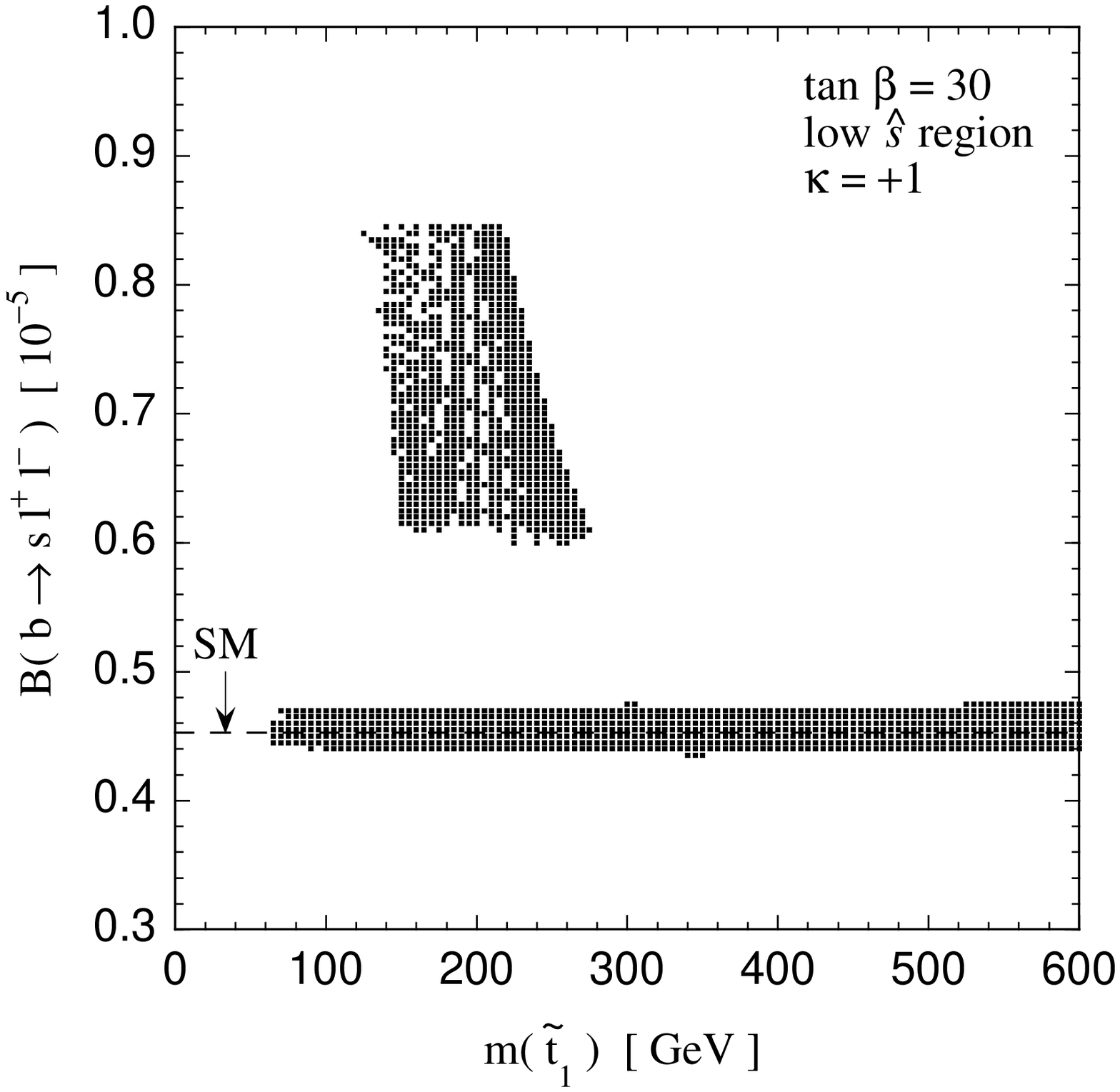}}
\\
    {\vskip 4cm}
    {\huge\bf Fig.~7 (b)}
  \end{center}
\end{figure}

\begin{figure}[p]
  \begin{center}
    \leavevmode
    \centerline{\psfig{file=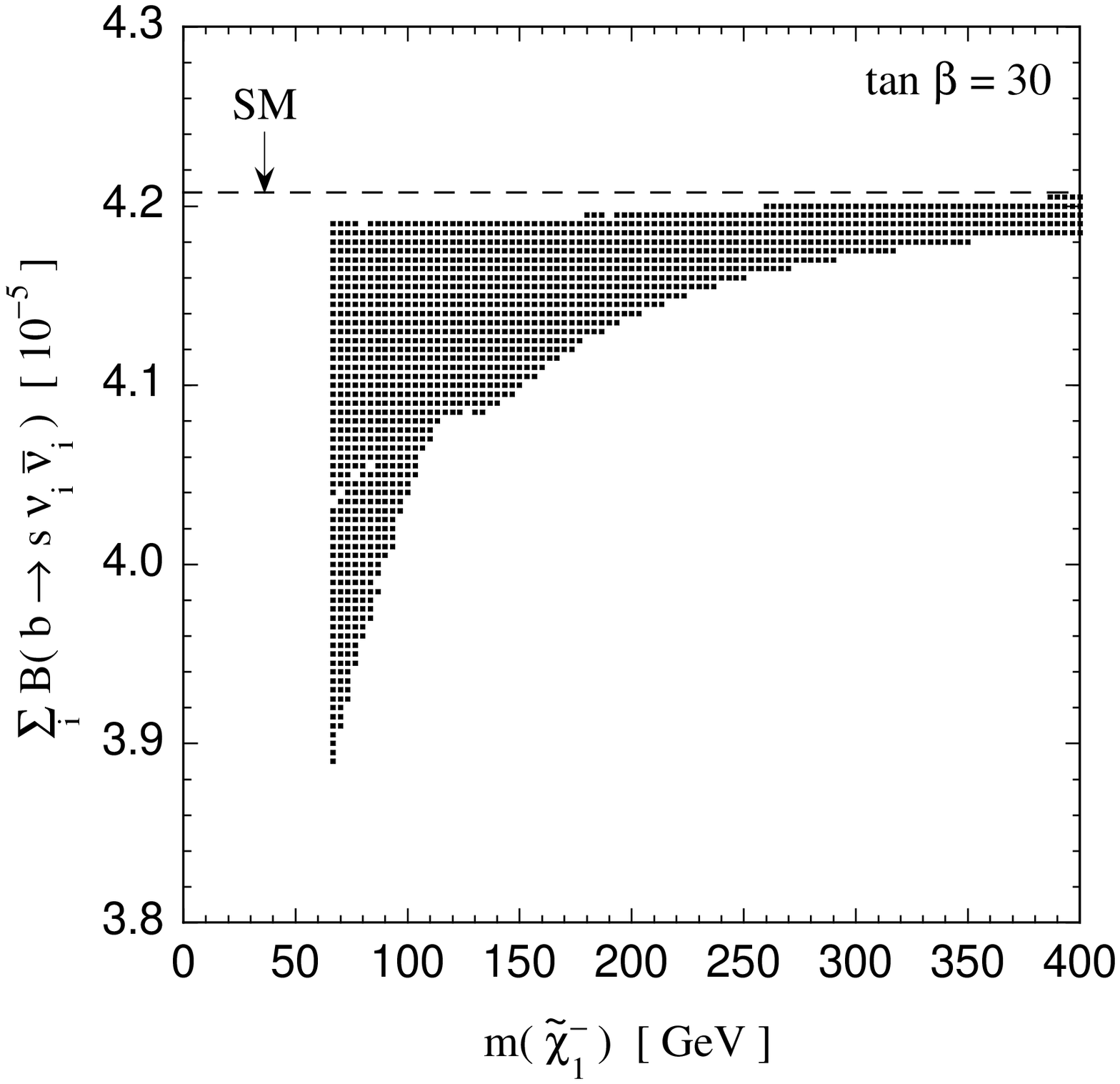}}
\\
    {\vskip 4cm}
    {\huge\bf Fig.~8 (a)}
  \end{center}
\end{figure}

\begin{figure}[p]
  \begin{center}
    \leavevmode
    \centerline{\psfig{file=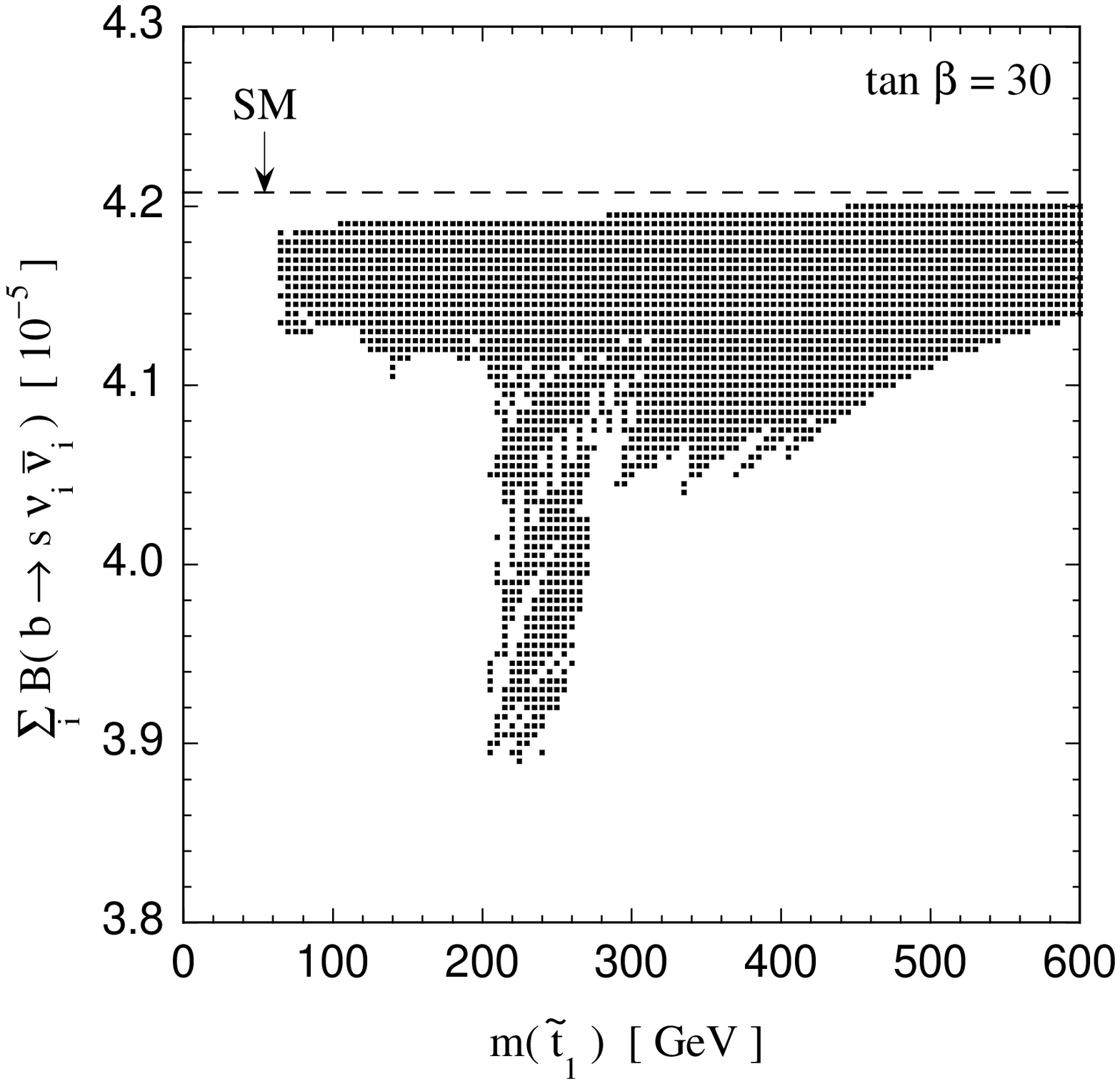}}
\\
    {\vskip 4cm}
    {\huge\bf Fig.~8 (b)}
  \end{center}
\end{figure}

\end{document}